\documentclass[11pt]{article}
\usepackage[utf8]{inputenc}
\usepackage{amsmath}
\usepackage{graphicx}
\usepackage{booktabs}
\usepackage{multirow}
\usepackage{url}
\usepackage{hyperref}
\usepackage{float}
\usepackage{caption}
\usepackage{subcaption}
\usepackage{amssymb}

\usepackage[top=2.5cm, bottom=2.5cm, left=2cm, right=2cm]{geometry}
\usepackage{setspace}
\title{Application of Deep Learning to Jet Charge Discrimination}
\author{Meisam Ghasemi Bostanabad and Mojtaba Mohammadi Najafabadi\\
School of Particles and Accelerators,\\
Institute for Research in Fundamental Sciences (IPM)\\
P.O. Box 19395-5531, Tehran, Iran}
\date{}

\begin{document}

\maketitle

\begin{abstract}
The Large Hadron Collider (LHC) produces an enormous volume of data in which the identification and characterization of hadronic jets is a central challenge. Determining the electric charge of the parton initiating a light-quark jet; a task known as jet-charge discrimination; is highly valuable for both precision tests of the Standard Model (SM) and searches for physics beyond it. In this work, we benchmark a range of classical and quantum machine-learning models for the task of distinguishing up-quark from anti-up-quark jets in a controlled QCD environment. Among the approaches tested, a Graph Neural Network achieved the best performance, with an AUC of 0.883. Jet-charge tagging of this kind has broad phenomenological applications, from improving measurements of charge asymmetries to enhancing sensitivity in searches for new particles from beyond the SM where quark versus antiquark discrimination is essential. Our study provides a methodological foundation for deploying modern machine-learning techniques in jet-charge analyses at the LHC experiments.
\end{abstract}

\section{Introduction}
The LHC is the world's most powerful and complex particle accelerator, designed to probe the fundamental laws of nature \cite{evans2008lhc}. Its primary goals include testing the predictions of the SM of particle physics, such as the existence of the Higgs boson, while also searching for signs of new physics beyond the SM, such as supersymmetry, two-Higgs-doublet models, and extra dimensions. However, one of the greatest challenges in analyzing the enormous volumes of data generated by the LHC is the study of hadronic jets; collimated sprays of particles resulting from quantum chromodynamic (QCD) processes. These jets are produced in high-energy collisions and their detailed characterization is essential. Accurate identification of the origin of these jets, whether from quarks or gluons \cite{krohn2013jet,gallicchio2013quark}, is critical for testing the predictions of the SM and for uncovering any potential signals of new, unknown phenomena. Improving our ability to use electric charge to discriminate between jets originating from different particles will enhance our understanding of both well-established theories and potential discoveries at the LHC.

\begin{figure}[htbp]
\centering
\includegraphics[width=0.35\linewidth]{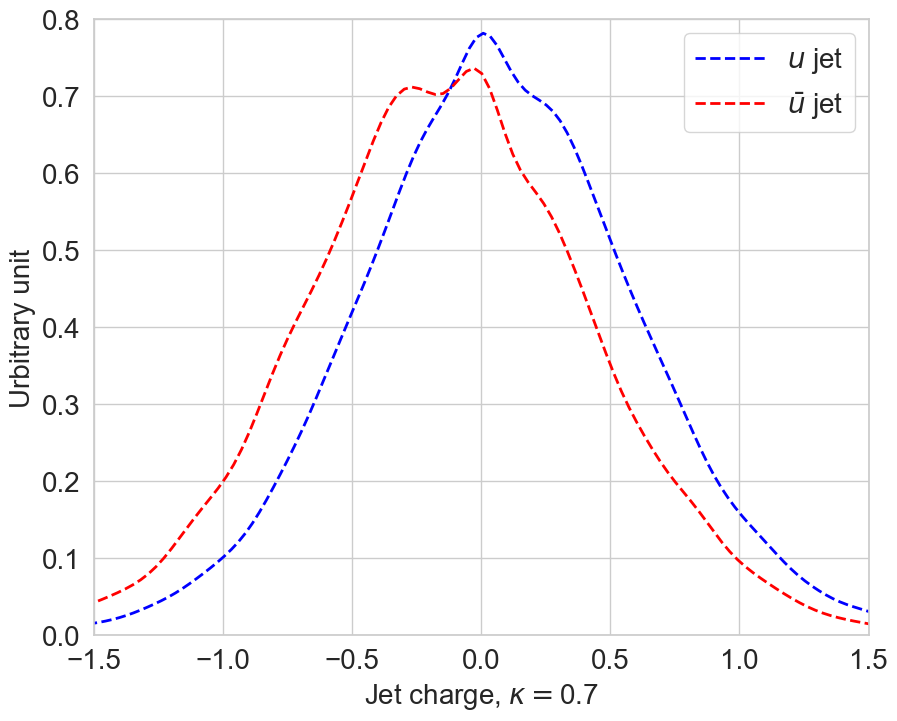}
\includegraphics[width=0.35\linewidth]{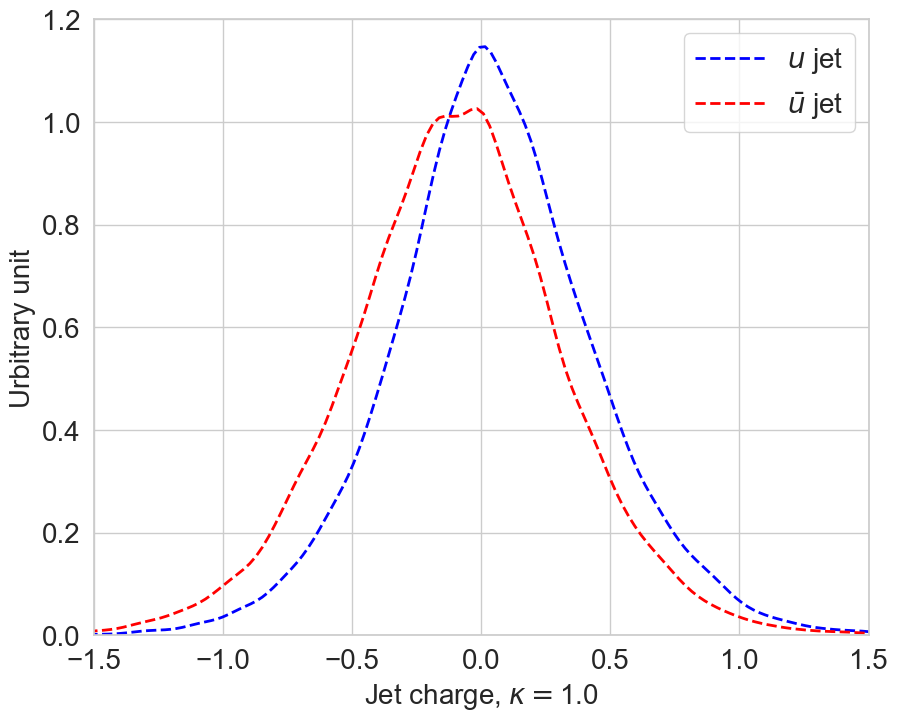}
\caption{Distributions of $\mathcal{Q}_{j}$ for $u,\bar{u}$ jets obtained from $pp\to ug$ or $pp\rightarrow\bar{u}g$ events with $\kappa=0.7,1$.}
\label{fig:jet_charge_dist}
\end{figure}

The concept of the jet charge observable, first introduced by Field and Feynman \cite{field1978parametrization}, has been explored in experimental studies such as measurement of top quark charge at the LHC \cite{atlas2013top}, and identifying the charge of heavy bosons $W^{\prime}/Z^{\prime}$ \cite{krohn2013jet}. Previous studies of jet charge \cite{cms2013measurement} involve measuring variants of a momentum-weighted jet charge, typically defined as a summation of the particle tracks within the jet, weighted by their transverse momentum ($p_{T}$) \cite{krohn2013jet}:

\begin{equation}
\mathcal{Q}_{j}=\frac{1}{(p_{T}^{\rm jet})^{\kappa}}\sum_{i\in\rm Tr}Q_{i}(p_{T}^{i})^{\kappa}
\label{eq1}
\end{equation}

where the summation runs over all particles tracks ($\rm Tr$) recorded for the jet, $Q_{i}$ and $p_{T}^{i}$ represent the charge of the object and the magnitude of its transverse momentum relative to the beam axis, and $p_{T}^{\rm jet}$ is the total transverse momentum of the jet. The parameter $\kappa$ is a tuning parameter defining the weighting with values ranging from 0 to 1. When $\kappa$ approaches 0, the jet charge becomes highly influenced by soft tracks, which are often not detectable. Conversely, as $\kappa$ increases towards 1, the jet charge becomes primarily determined by the charge of the leading track \cite{berge1981observation,waalewijn2012jet,aleph1991measurement}. Fig. \ref{fig:jet_charge_dist} presents the distributions of $\mathcal{Q}_{j}$ for $u,\bar{u}$ jets at two values of $\kappa$. Although there is a large overlap between the positive and negative jet charge distributions, $\mathcal{Q}_{j}$ remains useful for distinguishing the charge of the originating parton.

Recent developments, such as the proposal of the dynamic jet charge \cite{kang2021dynamic}, which generalizes the momentum-fraction weighting in the standard jet charge definition, have demonstrated enhanced discrimination power between quark and gluon jets as well as between different quark flavors, highlighting the potential of jet charge as a versatile tool in collider phenomenology.

Traditional jet charge methods \cite{kang2021dynamic,atlas2013top} typically use momentum-weighted sums of track charges, but their ability to distinguish between quark flavors is often limited by fluctuations and high correlation. In contrast, modern machine learning models particularly deep neural networks and graph neural networks learn directly from low-level features like particle momentum, position, and charge. These models are capable of capturing complex patterns and subtle relationships between jet constituents that handcrafted observables might miss, offering a promising path toward more accurate jet flavor identification. Indeed, machine learning techniques have already demonstrated impressive success \cite{cms2014measurement,ghasemi2025machine,cms2009particle} in the ATLAS \cite{atlas2020anomaly} and CMS \cite{cms2008experiment} experiments, where they have been widely adopted for tasks such as particle identification \cite{baldi2014searching1}, event reconstruction \cite{radovic2018machine}, and anomaly detection \cite{collins2018anomaly}.

This paper is organized as follows. A description of physics motivations is given in Section \ref{sec2}. Section \ref{sec3} describes the simulation setup and the input features used for training, including event generation, jet clustering, and variable construction. Section \ref{sec4} presents the performance of different machine-learning approaches, covering deep neural networks, convolutional neural networks, graph neural networks, quantum machine-learning models, as well as other classical classifiers, and also includes an extension to down and anti-down quark discrimination. In Section \ref{sec5}, we summarize the findings, discuss the implications of the results. Finally, section \ref{sec6} concludes the paper and highlights directions for future work.

\section{Physics Motivation and Scope}
\label{sec2}

The determination of the electric charge of jets, commonly referred to as jet-charge discrimination, is a powerful technique in collider physics. Although jets are extended objects formed through parton showering and hadronization, their overall charge encodes information about the initiating parton. This information has already proven useful in SM studies, for example, in determining the charge of the $W$ boson in hadronic decays, in probing the properties of the top quark, and in constraining parton distribution functions (PDFs) via charge asymmetries in jet production \cite{atlas2016new,cms2014identification}. Beyond the SM, jet charge provides a unique handle in processes where distinguishing quark from antiquark jets allows one to construct asymmetry observables sensitive to fundamental couplings or potential new sources of CP violation.

A particularly compelling case arises in the context of charged Higgs bosons ($H^{\pm}$), which are predicted in many extensions of the SM, including supersymmetry and two-Higgs-doublet models (2HDMs) \cite{branco2012theory}. At the LHC, production mechanisms such as $gb \to tH^{-}$ and $g\bar{b} \to \bar{t}H^{+}$ arise naturally in these scenarios. In such processes, the ability to determine the charge of final-state objects, in particular through jet-charge techniques in hadronic decays, is essential for distinguishing between $H^{+}$ and $H^{-}$ production and for constructing charge-asymmetry observables sensitive to the underlying couplings. Furthermore, the $tbH^{\pm}$ interaction is chiral, with couplings of the schematic form \cite{branco2012theory}

\begin{equation}
{\cal L}_{tbH^{\pm}}\propto\frac{g}{\sqrt{2}m_{W}}\left[m_{t}\cot\beta\,\bar{t}P_{L}b+m_{b}\tan\beta\,\bar{t}P_{R}b\right]H^{+}+{\rm h.c.},
\label{eq:lagrangian}
\end{equation}

Here, $m_{t}$, $m_{b}$, and $m_{W}$ denote the masses of the top quark, bottom quark, and $W$ boson, respectively. The left- and right-handed projections are denoted by $P_{L}$ and $P_{R}$, while $\tan\beta$ represents the ratio of the vacuum expectation values of the two Higgs doublets. The interplay between the left- and right-handed components can lead to observable differences in \textit{charge-asymmetry observables}, such as

\begin{equation}
A_{H}=\frac{N(H^{+})-N(H^{-})}{N(H^{+})+N(H^{-})},
\label{eq:asymmetry}
\end{equation}

which are sensitive to $\tan\beta$ and to possible CP-violating phases in the Higgs sector. Hadronic decays of the charged Higgs boson provide natural opportunities to exploit jet charge. For example, in the low-mass regime ($m_{H^{\pm}}<m_{t}$), the dominant decay channel can be $H^{\pm}\to c\bar{s}$, which produces two light-quark jets with opposite charge. The ability to correctly identify the charge of the charm jet (distinguishing $c$ from $\bar{c}$) is crucial for tagging whether the parent boson was $H^{+}$ or $H^{-}$. In this channel, jet-charge tagging can be used to build charge asymmetry observables directly from the dijet system, offering a complementary strategy to rate-based searches. Furthermore, since $H^{\pm}\to c\bar{s}$ suffers from large QCD backgrounds, improved jet-charge discrimination enhances sensitivity by both suppressing background and preserving the sign information of the signal. At higher masses, decays such as $H^{\pm}\to W^{\pm}h^{0}\to q\bar{q}^{\prime}\,b\bar{b}$ also benefit from jet-charge tagging, since the charge of the light-quark jet from the $W$ decay is correlated with the overall sign of the $H^{\pm}$. In these cases, precise jet-charge identification reduces combinatorial ambiguities in multi-jet final states, suppresses contamination from $W^{\pm}\to q\bar{q}^{\prime}$ and QCD multijet production, and enables charge-asymmetry measurements that probe both couplings and potential CP violation. Similar arguments apply to $H^{\pm}\to tb$ decays, where the charge of the $b$ jet and the decay products of the top quark provide complementary information.

In this work, however, our objective is not to carry out a phenomenological search for charged Higgs bosons or any specific BSM scenario. Instead, our aim is to develop and benchmark machine-learning algorithms for jet-charge discrimination in a controlled environment. To this end, we construct simplified QCD samples of processes such as $pp\to ug,\ pp\rightarrow\bar{u}g$, and related channels, in which a light quark or antiquark recoils against a gluon. These samples provide a well-defined laboratory: the initiating parton's charge is known unambiguously, the essential features of jet fragmentation and hadronization are preserved, and the complexity of multi-step decay chains is avoided. This setup allows us to isolate the algorithmic aspects of jet-charge discrimination and quantify performance in a transparent way. The insights gained here are directly transferable to more complex and phenomenologically relevant situations, including those involving charged Higgs bosons, where jet charge plays a decisive role.

\section{Simulation setup and input features}
\label{sec3}

The training dataset consists of simulated jet events, evenly divided between up quark and anti-up quark jets in the $pp\to ug$ and $pp\rightarrow\bar{u}g$ processes, generated with MadGraph5\_aMC@NLO\cite{alwall2014automated} and subsequently processed with PYTHIA8\cite{sjostrand2015introduction} for parton showering, hadronization, and underlying event modeling. For this study, jets are clustered using anti-$k_{t}$ algorithm \cite{cacciari2008anti} with radius parameter $R=0.4$, implemented through the FASTJET package \cite{cacciari2012fastjet,cacciari2006dispelling}. The PYTHIA generator is used to collect information for every jet in an event such as $p_{T}$, $\eta$ (the pseudorapidity, defined as $\eta=-\ln(\tan\frac{\theta}{2})$, where $\theta$ is the polar angle measured from the beam axis), $\phi$ (the azimuthal angle around the beam axis), momentum weighted charge for $\kappa$ of 0, 0.3, 0.5, 0.7, and 1. The jets are required to have a minimum transverse momentum of 10 GeV and a maximum absolute pseudorapidity of 2.5. These criteria ensure that only jets with sufficient energy and within a central detector region are considered for further analysis. Jet-level variables are used as inputs to the deep neural network. For the convolutional neural network, constituent-level information is provided, including each particle's $p_T$, $\eta$, $\phi$, and electric charge. Additionally, we incorporate observables motivated by traditional jet charge analyses \cite{rueter2015characterization}, defined as:

\begin{equation}
Q_{1,\kappa}=\sum_{i\in T_{T}}q_{i}z_{i}^{\kappa}
\label{eq:q1}
\end{equation}

\begin{equation}
Q_{2,\kappa}=\frac{\sum_{i\in T_{T}}q_{i}\,|\Delta\eta_{i}|^{\kappa}}{\sum_{i\in T_{T}}|\Delta\eta_{i}|^{\kappa}}
\label{eq:q2}
\end{equation}

where $z_{i}=p_{T_{i}}/p_{T_{jet}}$ is the fraction of the jet momentum, and the sum runs over every particle in a jet, which provides much more information than just the particles with leading $p_{T}$. The last variables used for model training are total jet charge as the sum of track's charge, and charge ratio as the sum of positive track charges divided by the absolute value of the sum of negative track charges:

\begin{equation}
\text{Charge Ratio}=\frac{\sum_{i\in Tr}q^{+}_{i}}{\left|\sum_{j\in Tr}q^{-}_{j}\right|}
\label{eq:charge_ratio}
\end{equation}

The complete set of input variables used in this study is summarized below:

\begin{itemize}
    \item Jet-level variables: transverse momentum $p_T$, pseudorapidity $\eta$, and azimuthal angle $\phi$.    
    \item Momentum-weighted jet charge observables $Q_{1,\kappa}$ and $Q_{2,\kappa}$ for $\kappa = 0, 0.3, 0.5, 0.7, 1$.    
    \item Total jet charge and charge ratio.      
    \item Constituent-level variables for each particle within a jet: transverse momentum $p_T$, pseudorapidity $\eta$, azimuthal angle $\phi$, and electric charge.
\end{itemize}

\section{Performance of Machine Learning Models}
\label{sec4}

In this section, we present the performance of different classical and quantum machine learning models in discriminating up-quark from anti-up-quark initiated jets. These models were trained using both high-level jet charge observables and low-level jet image representations constructed from charged particle constituents.

\subsection{Deep Neural Network}
A deep neural network (DNN) \cite{baldi2014searching} is a machine learning model inspired by the human brain, consisting of multiple layers of interconnected artificial neurons. Each neuron in the network performs a computation based on its inputs, weights, and biases. The relationship between the inputs and the output of a neuron can be expressed as $z=\sum_{i=1}^{n}w_{i}x_{i}+b$, where $w_{i}$ are the weights associated with each input $x_{i}$, and $b$ is the bias. The output $a$ of a neuron is obtained by applying a non-linear activation function $\sigma$, such that $a=\sigma(z)$. Weights and biases are crucial as they determine how the input data is transformed as it propagates through the network. Learning these parameters effectively allows the model to recognize patterns and make accurate predictions.

The training of a DNN involves forward and backward propagation. During forward propagation, the input data is passed through the network layer by layer, and each layer's output becomes the input for the next layer, ultimately generating a prediction. This process can be expressed as ${\bf a}^{(l+1)}=\sigma({\bf W}^{(l)}{\bf a}^{(l)}+{\bf b}^{(l)})$, where ${\bf W}^{(l)}$ represents the weights, ${\bf b}^{(l)}$ represents the biases of layer $l$, and ${\bf a}^{(l)}$ is the activation. In the backward propagation, the goal is to minimize the error between the predicted output and the actual output using an error function, such as mean squared error or cross-entropy. Gradients of the error function with respect to each parameter are calculated using the chain rule of differentiation. These gradients are then used in an optimization algorithm, such as gradient descent, to iteratively update the weights and biases, following the equation ${\bf W}^{(l)}\leftarrow {\bf W}^{(l)}-\eta\frac{\partial L}{\partial {\bf W}^{(l)}}$, where $\eta$ is the learning rate, and $L$ is the loss function. This process allows the network to improve its performance through multiple iterations and effectively learn the complex relationships within the data.

The DNN used in this study is a fully connected feedforward architecture designed for binary classification of up and anti-up quark jets. The model consists of three hidden layers, each containing 32 neurons with the hyperbolic tangent (tanh) activation function and weights initialized using a normal distribution. To prevent overfitting, dropout layers with a rate of 0.2 follow each dense layer. The final output layer is a single neuron with a sigmoid activation function, suitable for probabilistic binary output. The model is compiled using the stochastic gradient descent (SGD) optimizer and trained with the binary cross-entropy loss function. Prior to training, the input features were standardized using a StandardScaler method, ensuring that each feature had zero mean and unit variance to improve the convergence and stability of the DNN.

Hyperparameter optimization is performed using the RandomizedSearchCV method from the scikit-learn \cite{pedregosa2011scikit} library, wrapped around the Keras \cite{chollet2015keras} model via SciKeras \cite{nicolaou2021scikeras}. The search space includes variations in the number of layers, number of neurons per layer, dropout rate, activation functions, kernel initializers, batch size, and optimizer type. The search uses 5-fold cross-validation to evaluate model configurations and identify the one with the highest average accuracy on the training set. This strategy enables efficient tuning over a wide space of architectures and training settings while minimizing overfitting and improving generalization.

\begin{figure}[htbp]
\centering
\includegraphics[width=0.36\linewidth]{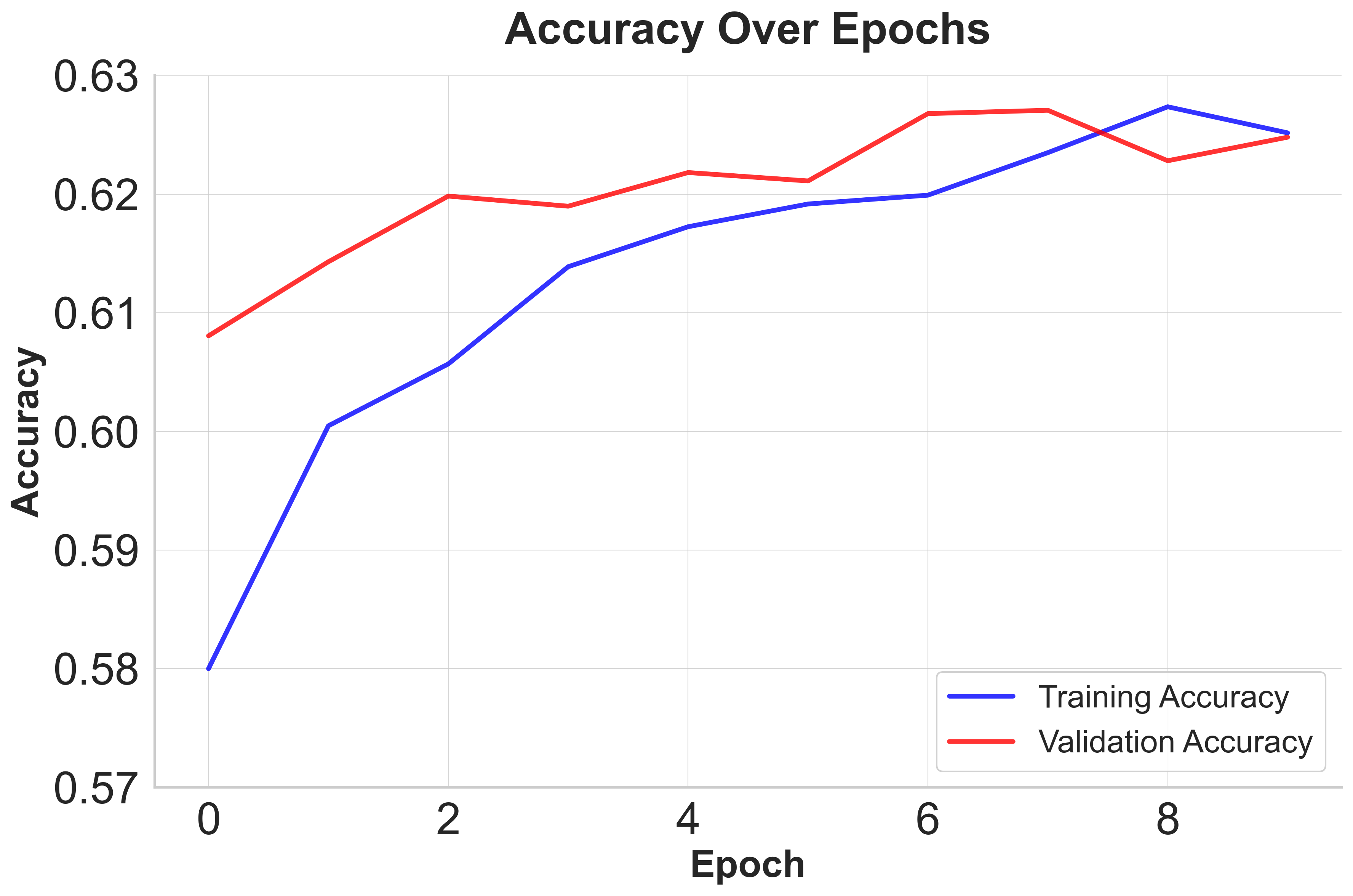}
\includegraphics[width=0.36\linewidth]{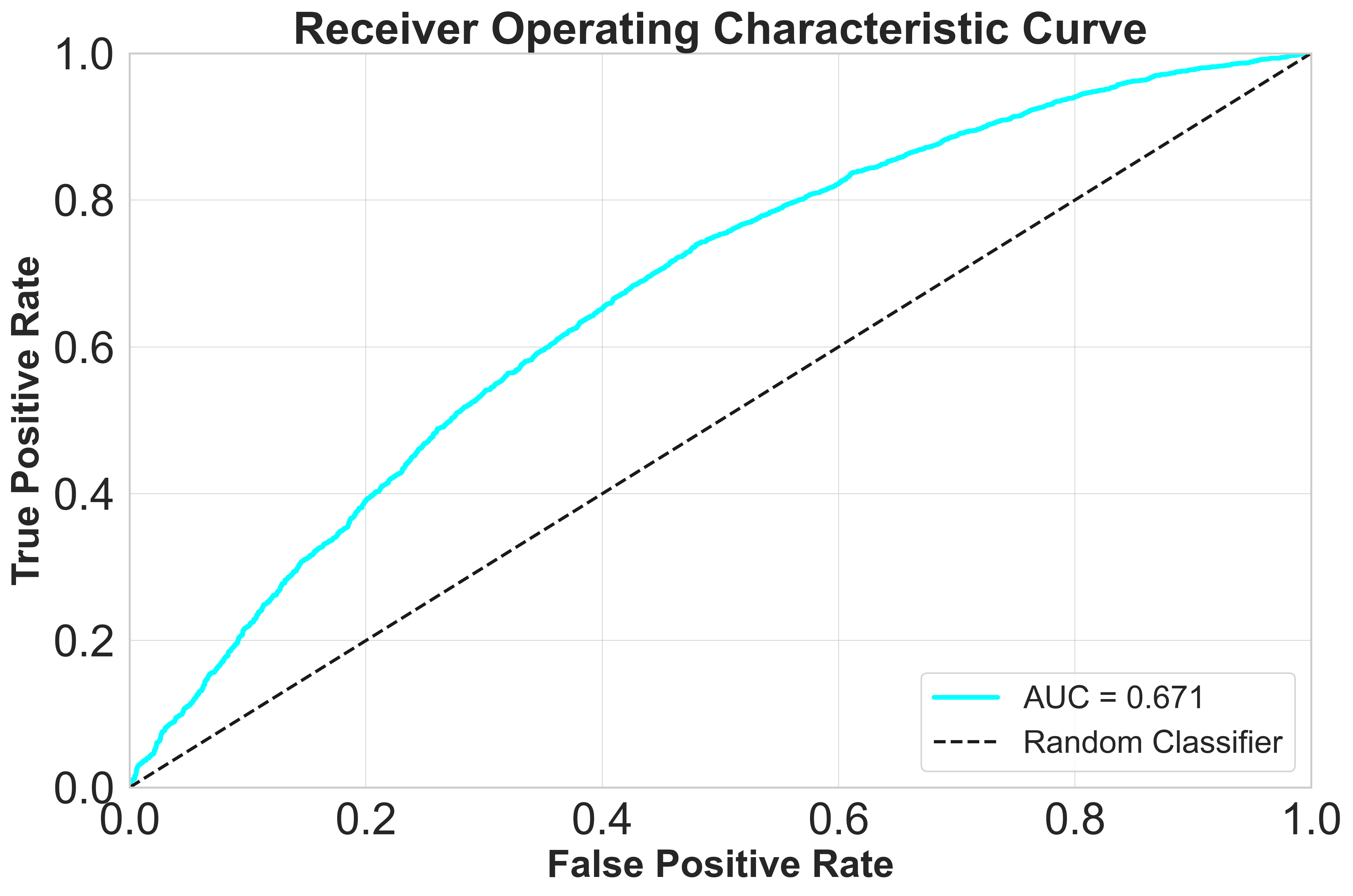}
\caption{Left: the accuracy progression over 16 training epochs, showing both training (blue) and validation (red) accuracy. Right: the Receiver Operating Characteristic (ROC) curve, with an Area Under the Curve (AUC) score of $0.671$, which measures the model's ability to distinguish between the two classes.}
\label{fig:dnn_results}
\end{figure}

Fig. \ref{fig:dnn_results} left shows the accuracy of the deep learning model over training epochs, where training (validation ) accuracy improves rapidly in the initial epochs before converging around $63\%$ ($62.5\%$). The relatively close alignment between training and validation accuracy suggests that the model generalizes reasonably well without significant overfitting. Fig. \ref{fig:dnn_results} right presents the receiver operating characteristic (ROC) curve for the model, with an area under the curve (AUC) of $0.671$, indicating that the model has moderate discriminatory ability in distinguishing between up and anti-up quark jets, with values closer to 1.0 indicating the best separation. The model outperforms random classification (AUC=0.5), represented by the dashed line, highlighting its ability to identify meaningful patterns in the data. Compared to more advanced architectures, the fully connected DNN was limited in capturing the complex spatial and relational dependencies within jets, making it less effective.

\subsection{Convolutional Neural Network}
A convolutional neural network (CNN) \cite{lecun1998gradient} is an advanced type of deep learning model particularly well-suited for tasks involving spatial data, such as image recognition. It consists of several key layers, including convolutional, pooling, and fully connected layers. The convolutional layers apply filters to the input data to learn spatial features, allowing the network to recognize complex patterns by combining simple ones from previous layers. The pooling layers help to reduce the size of the data representation, making the model more computationally efficient and less prone to overfitting. Finally, the fully connected layers use the extracted features for classification. By stacking multiple convolutional layers, a CNN learns to represent increasingly abstract and complex features from the raw input, ultimately making accurate predictions.

\begin{figure}[htbp]
\centering
\includegraphics[width=0.65\linewidth]{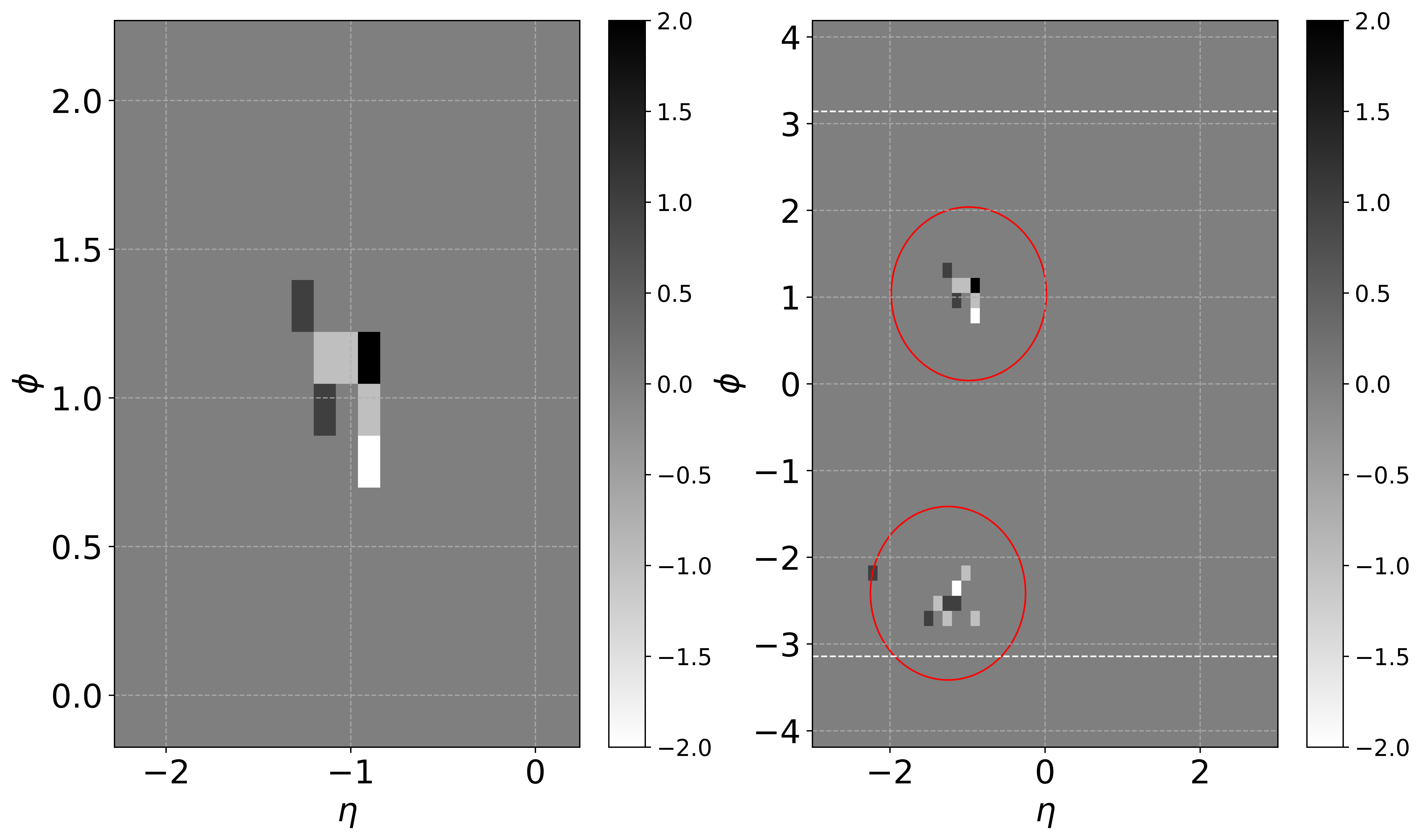}
\caption{Left: pixelated representation of a leading jet in the $\eta$-$\phi$ space, showing regions normalized by electric charge, used as input for the CNN model. Right: same representation for all jet's constituents in an event.}
\label{fig:cnn_input}
\end{figure}

In this case, the CNN is used to discriminate between up and anti-up quark jets by analyzing pixel charge information plotted in $\eta$ (pseudorapidity) and $\phi$ (azimuthal angle) space. The input to the CNN consists of images like the Fig. \ref{fig:cnn_input}, where each pixel represents a region in the $\eta$-$\phi$ space and is normalized by the electric charge of particles observed in that region. The electric charge of jet constituents is encoded directly in the pixel intensities of the input image, allowing the CNN to learn charge-dependent spatial patterns. The convolutional layers extract features from the spatial distribution of the pixel charge, identifying distinctive charge patterns that differentiate up and anti-up quarks. By recognizing these spatial features in the $\eta$-$\phi$ distribution, the CNN is capable of learning the inherent differences between the two quark types and effectively classifying them based on their charge characteristics.

The baseline CNN architecture designed for this study processes jet images formatted as $16\times 22$ charge-weighted ($\eta,\phi$) pixel maps. The model begins with a convolutional layer of 32 filters with a $3\times 3$ kernel and ReLU activation (defined as $f(x)=\max(0,x)$, introduces non-linearity to enhance the model's ability to capture complex patterns), followed by batch normalization, max pooling, and a 30\% dropout layer for regularization. A second convolutional block with 64 filters, batch normalization, and similar pooling and dropout layers is added to deepen the representation. The output is then flattened and passed through a dense layer with 64 neurons, followed by batch normalization and a higher dropout rate of 40\%. The final output layer is a single neuron with sigmoid activation, suitable for binary classification. The model is trained using the Adam optimizer \cite{kingma2014adam} with a learning rate of $10^{-4}$ and binary cross-entropy loss.

To improve performance, an optimized CNN was constructed using the Keras Tuner framework \cite{omalley2019keras} with randomized hyperparameter search. This version dynamically adjusts the number of filters in each convolutional layer (ranging from 32 to 128), dropout rates (between 20\% and 60\%), dense layer size (64 to 128 units), and the learning rate (sampled logarithmically between ($10^{-5}$ and $10^{-3}$)). A learning rate scheduler (ReduceLROnPlateau) was also used to reduce the learning rate when validation performance plateaued. The optimized model selected by the tuner achieved improved generalization by exploring a broader architectural and training configuration space, demonstrating the advantage of systematic hyperparameter optimization over fixed-structure models. Table \ref{tab:cnn_comparison} highlights the key architectural differences between the baseline and optimized CNN models, illustrating how systematic hyperparameter tuning led to improved generalization and performance.

\begin{table}[htbp]
\centering
\caption{Comparison of baseline and optimized CNN hyperparameters}
\begin{tabular}{|l|c|c|}
\hline
\textbf{Hyperparameter} & \textbf{Baseline CNN} & \textbf{Optimized CNN} \\
\hline
Conv1 filters & 32 & 128 \\
Conv2 filters & 64 & 128 \\
Kernel size & $5\times 5$ & $3\times 3$ \\
Activation & ReLU & ReLU \\
Dropout (after Conv1) & 30\% & 30\% \\
Dropout (after Dense) & 40\% & 30\% \\
Dense layer size & 64 & 128 \\
Optimizer & Adam & Adam \\
Learning rate & $10^{-4}$ & $10^{-5}$-$10^{-3}$ \\
Learning rate scheduler & None & LROnPlateau \\
Loss function & Binary entropy & Binary entropy \\
\hline
\end{tabular}
\label{tab:cnn_comparison}
\end{table}

\begin{figure}[htbp]
\centering
\includegraphics[width=0.5\linewidth]{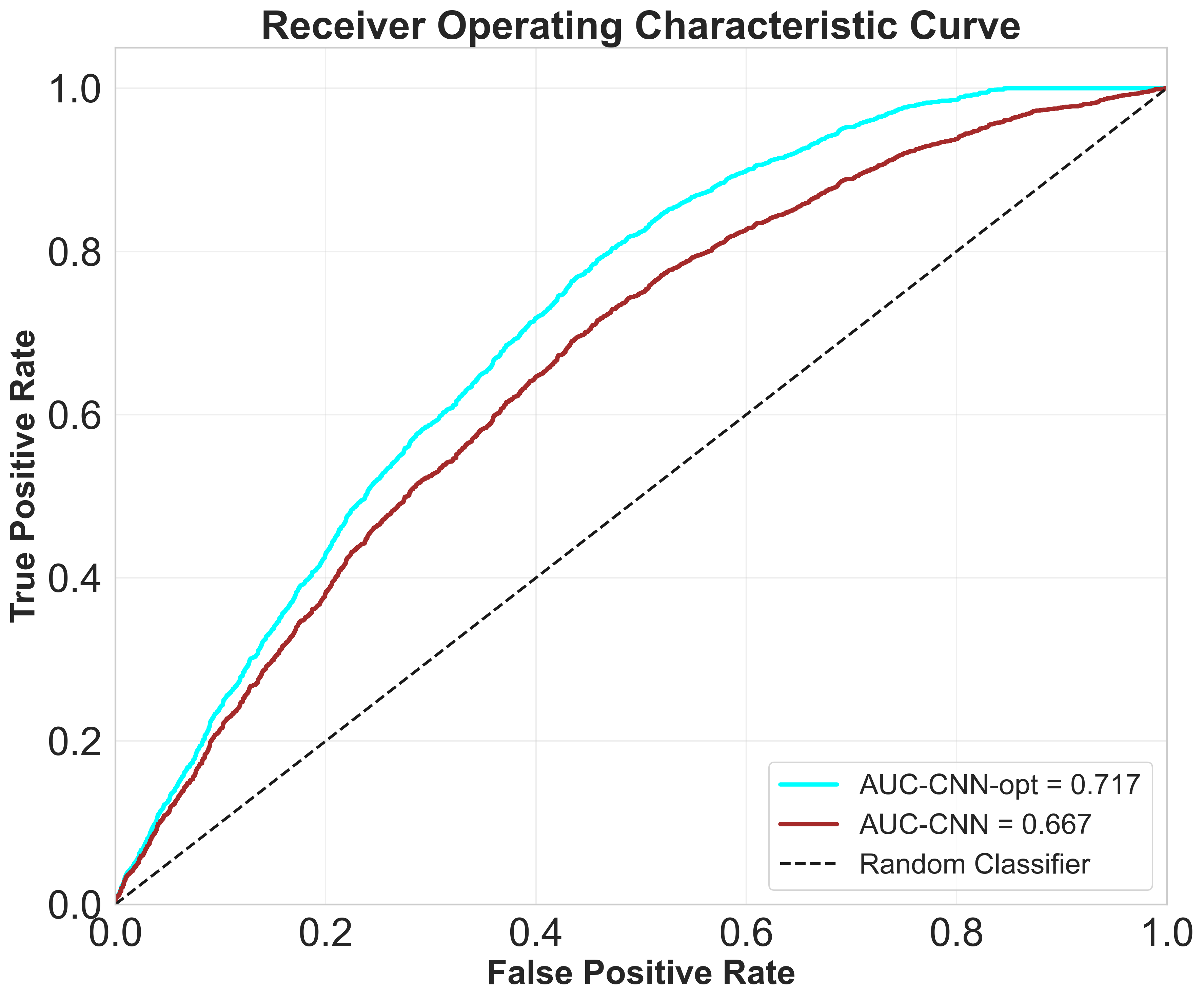}
\caption{ROC curves for baseline and optimized CNN models. The optimized model (AUC = 0.717) outperforms the baseline (AUC = 0.667), highlighting the benefit of hyperparameter tuning.}
\label{fig:cnn_roc}
\end{figure}

In Fig. \ref{fig:cnn_roc} the ROC curves comparing the baseline and optimized CNN models reveal a clear improvement in performance following the optimization process. The baseline model yields an AUC of 0.667, reflecting limited discrimination power. In contrast, the optimized CNN, obtained through a systematic hyperparameter search using Keras Tuner, achieves a substantially higher AUC of 0.717. This improvement underscores the importance of careful model design and tuning when working with jet image data. By adjusting key architectural parameters such as convolutional filter counts, dense layer sizes, dropout rates, and learning rates, the optimization process enhances the model's ability to generalize and capture subtle patterns that differentiate up from anti-up quark jets.

\subsection{Graph Neural Network}
Graph Neural Networks (GNNs) \cite{zhou2020graph} provide a powerful framework for analyzing data with an underlying graph structure. Unlike CNNs, which exploit local spatial correlations through convolutional filters defined on a fixed grid structure, GNNs are designed to capture pairwise relationships and interactions between all connected elements. This makes the GNN particularly well-suited for jet classification tasks, where particles within a jet are inherently linked through their spatial and kinematic correlations. By leveraging these connections, GNNs can model the internal structure of jets more effectively than standard architectures.

\begin{figure}[htbp]
\centering
\includegraphics[width=0.5\linewidth]{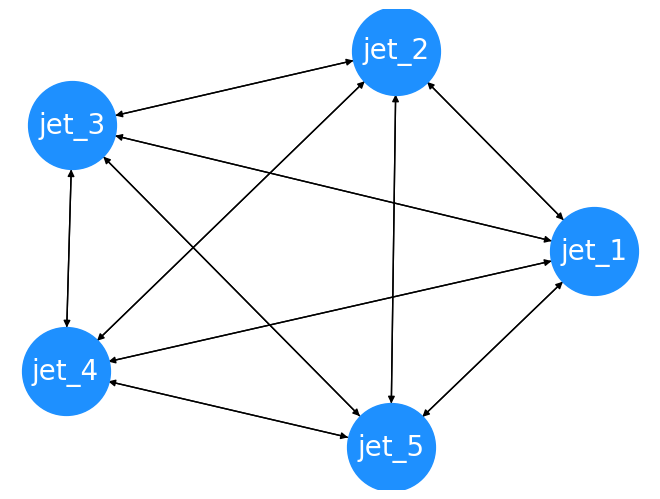}
\caption{Visualization of a fully connected directed graph used to represent jet constituents in the GNN model. Each node corresponds to a particle, and edges encode pairwise interactions without self-loops.}
\label{fig:gnn_graph}
\end{figure}

In our study, the jets are represented as fully connected graphs, where each particle within a jet corresponds to a node, and edges represent the relationships between particles. The node features include physical attributes such as transverse momentum, pseudorapidity, azimuthal angle, and momentum-weighted electric charge. The graph connectivity is defined using an edge index tensor, which lists all directed connections between node pairs. In our initial setup, we use a fully connected graph, meaning each node is linked to every other node, except itself, as shown in Fig. \ref{fig:gnn_graph}. While this approach simplifies the implementation, it can be refined by incorporating more physically motivated connections--for example, linking particles based on their spatial proximity or other shared properties relevant to jet structure.

Two graph neural network architectures Graph Convolutional Network (GCN) and GraphSAGE are employed for the task of graph classification, aiming to distinguish between up and anti-up quark jets. Graph Convolutional Networks (GCNs) \cite{kipf2017semi} are a natural extension of traditional convolutional networks to data with an underlying graph structure. Unlike standard neural networks that assume a fixed grid, GCNs are designed to work with data where elements are connected in more complex ways such as the particles within a jet. In a GCN, each node learns by gathering information from its neighboring nodes through a process known as message passing. This allows the model to capture not only the individual properties of particles but also how they relate to one another. With each layer, the network refines these representations by aggregating and transforming the input from surrounding nodes. As a result, GCNs are able to learn meaningful patterns from the graph's structure and are well-suited for tasks like classifying entire jets based on the relational information between their constituents.

The network consists of two graph convolutional layers, each followed by ReLU activation function and dropout layers with a dropout rate of 0.5 to mitigate overfitting. A global mean pooling operation is used to aggregate node features into a fixed-size vector representing the entire graph. This vector is then passed through a fully connected layer to produce the final classification output. The model is trained using the Adam optimizer with a learning rate of 0.001 and weight decay of $10^{-3}$, minimizing the cross-entropy loss. Training is performed over multiple epochs with early stopping based on test accuracy, and the model achieving the highest validation performance is saved for evaluation. This setup enables the GCN to learn from the relational structure of jet constituents and effectively distinguish between quark flavors.

GraphSAGE \cite{hamilton2017inductive} (Graph Sample and Aggregate) is a graph neural network architecture designed to efficiently learn node and graph representations by aggregating information from local neighborhoods using a learnable function in this case, a simple mean aggregator. Unlike traditional GCNs, which rely on the full adjacency structure, GraphSAGE samples a fixed-size set of neighbors and can generalize to unseen graphs, making it well-suited for jet-level classification tasks. In this analysis, the model architecture includes two SAGEConv layers with ReLU activations, each followed by dropout to mitigate overfitting. Node embeddings are aggregated using global mean pooling and passed through a fully connected layer to perform binary classification of up versus anti-up quark jets. The model is trained using the Adam optimizer with a learning rate of 0.001 and weight decay, using cross-entropy loss and early stopping to select the best-performing configuration. This setup allows the model to capture the structural and relational information within each jet graph while maintaining strong generalization performance.

\begin{figure}[htbp]
\centering
\includegraphics[width=0.4\linewidth]{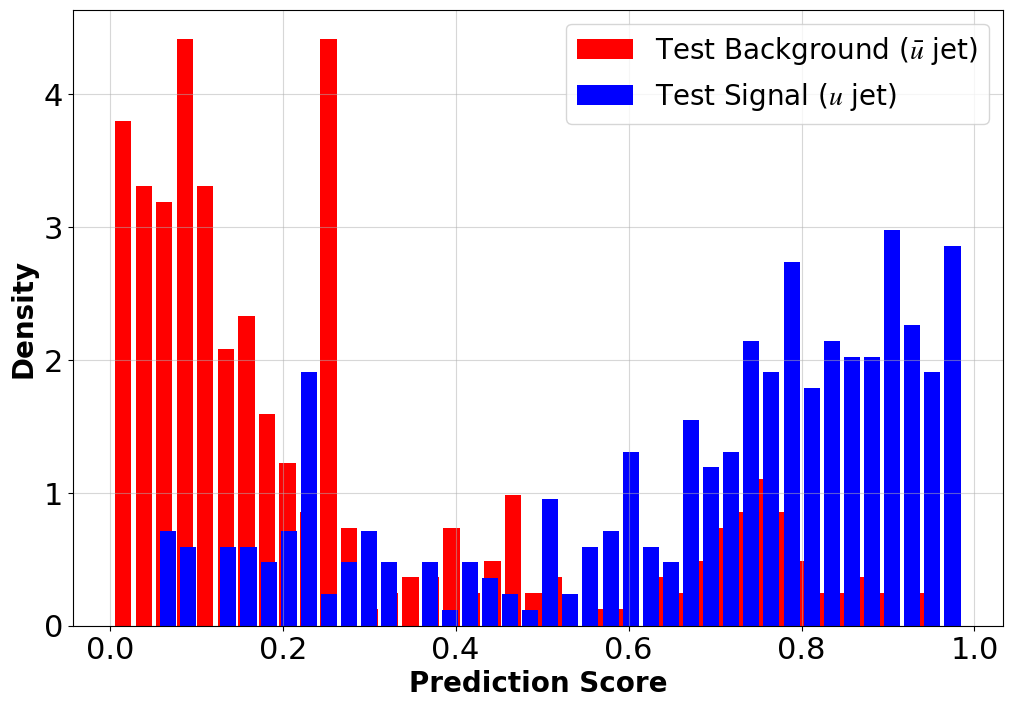}
\includegraphics[width=0.43\linewidth]{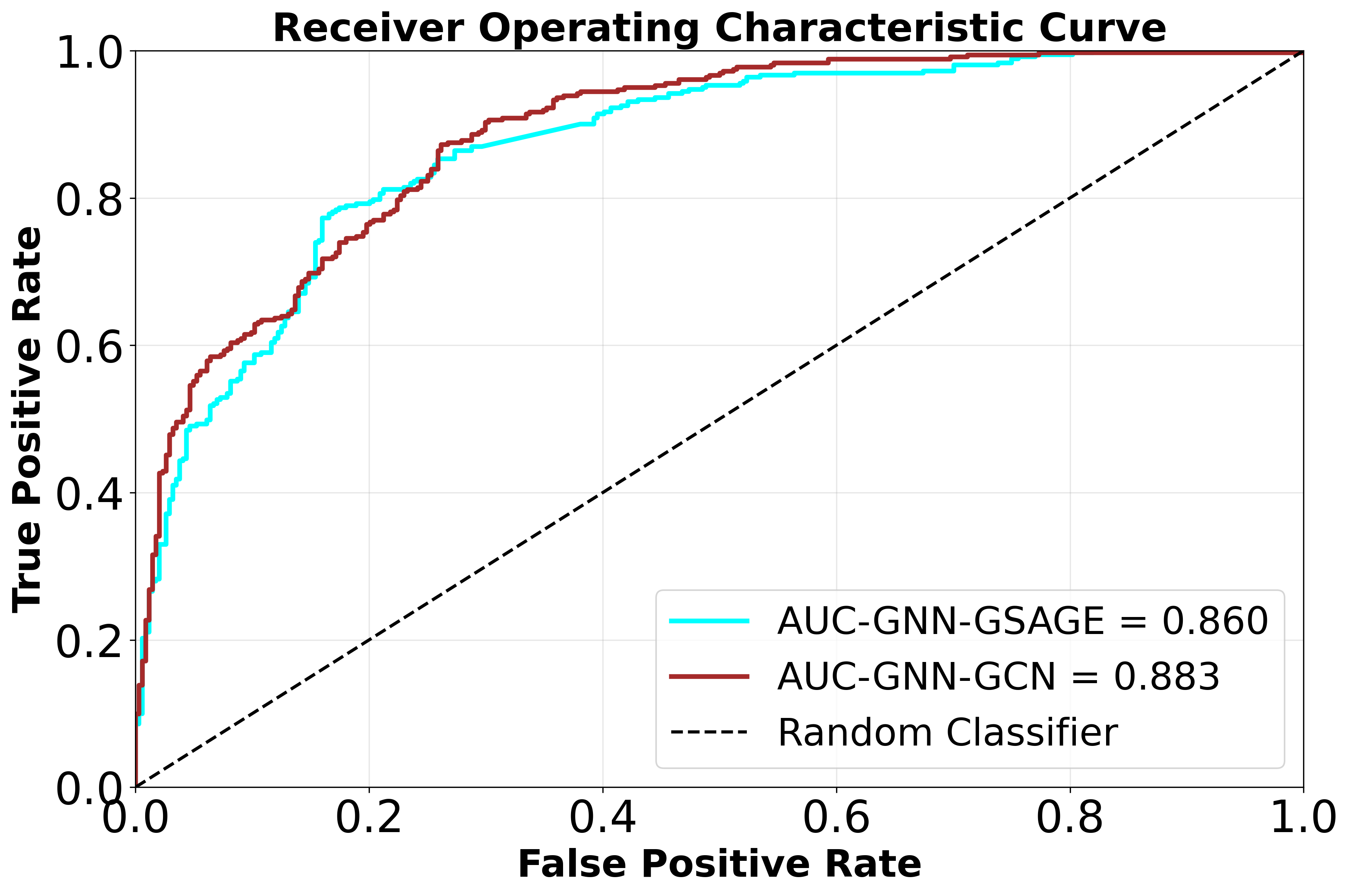}
\caption{Left: Distribution of GCN output scores for up (signal) and anti-up (background) jets, showing clear separation between the two classes. Right: ROC curves comparing GCN and GraphSAGE performance on jet classification.}
\label{fig:gnn_results}
\end{figure}

Fig. \ref{fig:gnn_results} (left) presents the output score distribution from the GCN model for up (signal) and anti-up (background) quark jets. The two distributions are clearly separated, with up jets peaking closer to 1 and anti-up jets clustering near 0. This strong separation shows that the model has learned to distinguish between the two jet types effectively based on their internal structure and charge-related patterns. The ROC curves in the Fig. \ref{fig:gnn_results} (right) shows that both Graph Neural Network models (GCN and GraphSAGE) achieve strong classification performance in distinguishing up from anti-up quark jets. The GCN slightly outperforms GraphSAGE, with an AUC of 0.883 compared to 0.860, indicating that while both models effectively capture relational information within jet graphs, the GCN provides a marginally better overall separation in this task. This demonstrates their strong suitability for jet charge discrimination tasks, especially in cases where subtle topological information is critical. Because we use fully connected jet graphs, runtime and memory scale quadratically with the number of constituents, which explains the higher training cost we observe relative to DNN/CNN.

\subsection{Transformer-based models}
Transformers have emerged as a powerful class of deep learning models built around the attention mechanism, originally introduced for sequence modeling by Vaswani et al. \cite{vaswani2017attention} in the "Attention Is All You Need" paper, and now widely used for set-structured data. In a Transformer, each input element (token) is mapped to a latent representation and is iteratively updated by self-attention, where the model learns data-driven weights that determine how strongly each token should aggregate information from all other tokens. This mechanism is particularly suitable for jets, since jets are naturally described as unordered collections of constituents (particles), where the relevant discriminating information arises from correlations across constituents. In jet tagging, Transformer models have demonstrated competitive performance by operating directly on constituent-level information and learning global dependencies through attention \cite{qu2022particletransformer,mikuni2021particleformer}.

In this work we implement a plain Particle Transformer to discriminate up versus anti-up-quark initiated jets. Unlike the full Particle Transformer proposed in \cite{qu2022particletransformer}, where attention logits are augmented by a learned pairwise “interaction” term, our implementation uses no interaction embedding. We use standard multi-head self-attention without adding pairwise biases and positional embedding, yielding a clean baseline that isolates what can be learned from tokent embeddings alone. This configuration allows a direct and controlled comparison with the DNN, CNN. and GNN baselines discussed previously.

\begin{figure}[htbp]
\centering
\includegraphics[width=\linewidth]{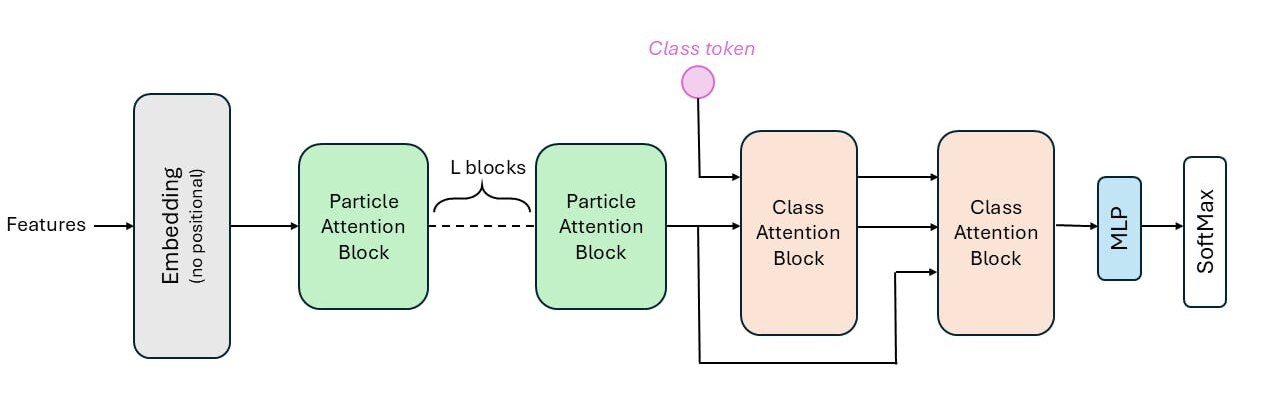}
\caption{The architecture of the Particle Transformer includes Embedding layer, Particle Attention Block and Class Attention Block.}
\label{fig:trans_struct}
\end{figure}

The model pipeline (illustrated in Fig. \ref{fig:trans_struct}) consists of three stages: (i) Particle Embedding: Each jet is represented by a fixed set of input features. These inputs are mapped into a latent space of dimension C through a small multilayer perceptron (Embedding block), producing a sequence of token embeddings $x^{0} \in \mathbb{R}^{P \times d}$, where $P$ denotes the number of tokens and $d$ represents the embedding dimension. (ii) Particle Attention Blocks (L blocks): The embedded tokens are processed by a stack of L Transformer blocks, each composed of multi-head self-attention with residual connections and layer normalization. Self-attention updates every token by aggregating information from all other tokens, enabling the model to learn global charge and kinematic patterns relevant for quark–antiquark discrimination. Formally, for input embeddings X, each block computes: $\mathrm{Attention}(Q,K,V)=\mathrm{Softmax}\!\left(\frac{QK^{\top}}{\sqrt{d_k}}\right)V$, where Q, K, and V are learned linear projections of X. (iii) Class Attention Blocks and Classification Head: After the particle-attention stack, a learnable class token is introduced and refined via a small number of Class Attention Blocks. In these blocks, only the class token is updated by attending to the particle tokens, producing a compact jet-level representation. This
representation is passed through an MLP classifier and a final softmax layer to obtain the probability of the jet being initiated by a up or anti-up quark.

The Transformer is trained for binary classification using cross-entropy loss. The architecture employs $L = 8$ particle attention blocks and $2$ class attention blocks, with an embedding dimension $C = 128$ and $8$ attention heads per block. Each feed-forward network expands the latent dimension by a factor of $4$, corresponding to a hidden size of $512$. Dropout with rate $0.1$ is applied within particle attention blocks, while class attention blocks are run without dropout to stabilize the global representation learning. The model is optimized using the AdamW optimizer with a learning rate of $2 \times 10^{-4}$ and weight decay of $1 \times 10^{-4}$. Training is performed with a batch size of $512$ for multiple epochs on a CUDA-enabled GPU, optionally using automatic mixed precision (AMP) to accelerate matrix multiplications while maintaining numerical stability. 

\begin{figure}[htbp]
\centering
\includegraphics[width=0.5\linewidth]{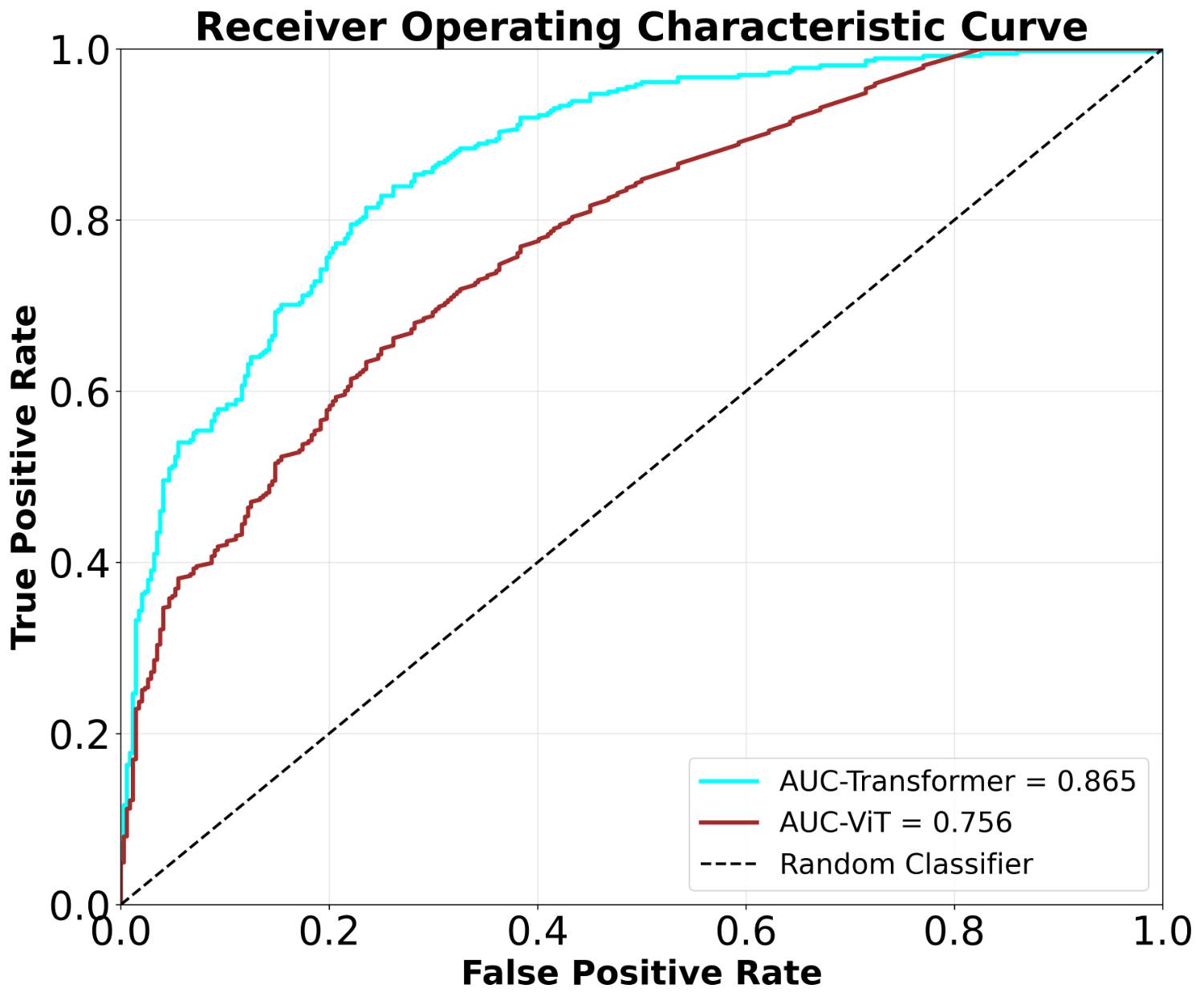}
\caption{ROC curve for the plain Particle Transformer. The model achieves an AUC of 0.865, demonstrating strong separation between the two classes across a wide range of operating thresholds.}
\label{fig:trans_roc}
\end{figure}

The ROC curve for the plain Particle Transformer is shown in Fig. \ref{fig:trans_roc}, yielding an AUC of 0.865. The curve rises steeply at low false-positive rates, indicating that the model is able to correctly identify a large fraction of up-initiated jets while maintaining strong background rejection.  The smooth approach of the ROC curve toward unity at high signal efficiency further indicates stable probabilistic separation across a wide range of operating thresholds, confirming that the learned representation provides meaningful discrimination power over the full decision spectrum.

\begin{figure}[htbp]
\centering
\includegraphics[width=0.65\linewidth]{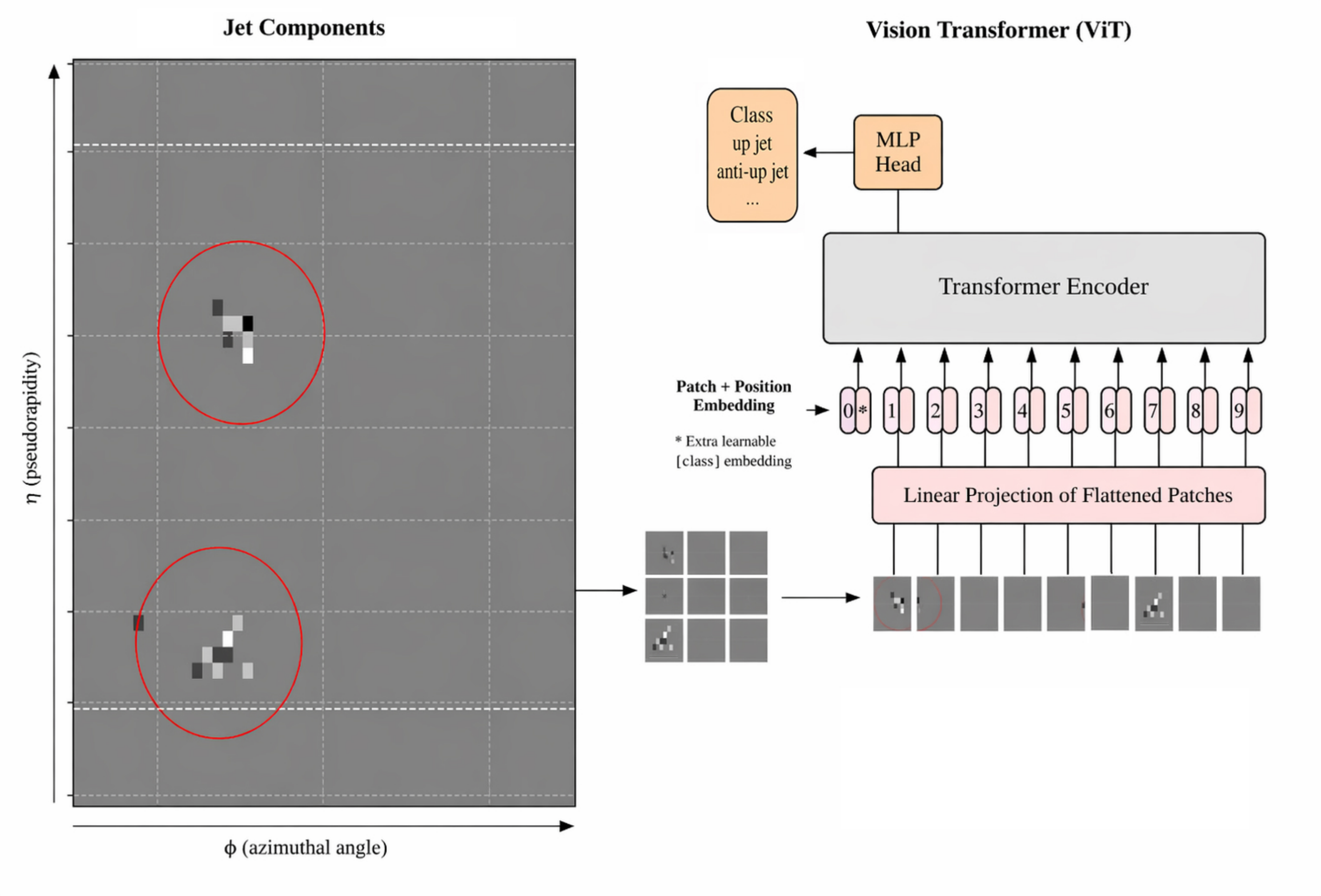}
\caption{Vision Transformer architecture for jet charge discrimination, showing patch embedding, transformer encoder blocks, and classification head.}
\label{fig:ViT}
\end{figure}

In addition to the standard transformer architecture, we also evaluate the Vision Transformer (ViT) \cite{dosovitskiy2021vit}, which applies the transformer mechanism to image classification by treating an image as a sequence of patches processed through self-attention. For our jet images of shape $(16 \times 22)$, each patch is flattened and linearly projected into an embedding space of dimension 64, with positional embeddings added to retain spatial information. A learnable class token is prepended to the sequence to serve as the aggregate representation for classification. Our ViT implementation consists of 4 transformer encoder blocks, each containing multi-head self-attention (4 heads) and feed-forward layers with residual connections and layer normalization. The multi-head attention mechanism computes attention across all patch pairs, enabling the model to capture long-range spatial dependencies in jet energy deposits. After the transformer blocks, the class token output passes through a dense layer with sigmoid activation for binary classification. The model is trained using the Adam optimizer and binary cross-entropy loss with dropout rate 0.25, consistent with the other architectures in this study. The architecture of the ViT includes patch embedding with position encoding, transformer encoder blocks, and an MLP classification head for up vs. anti-up jet discrimination is shown in Fig. \ref{fig:ViT}.

The performance of the ViT model is shown in Fig. \ref{fig:trans_roc}. While both the optimized CNN and the ViT share an identical input representation, the ViT achieves a superior AUC of $0.756$ relative to $0.717$ obtained by the optimized CNN, indicating that the self-attention mechanism, by capturing global spatial correlations among all patch pairs simultaneously, provides a more expressive representation of the charge distribution within the jet image than the locally constrained receptive fields of convolutional filters. 

\subsection{Quantum Machine Learning Models}
\begin{figure}[htbp]
\centering
\includegraphics[width=0.43\linewidth]{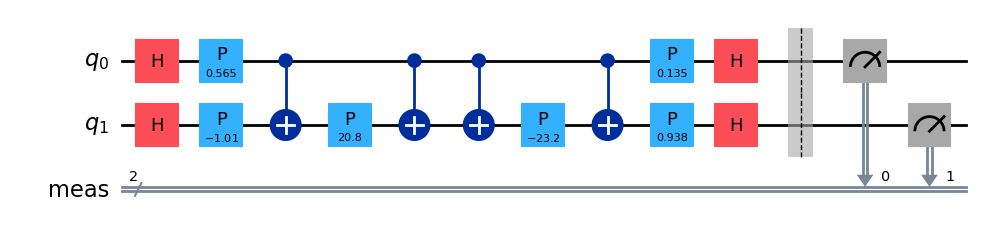}
\includegraphics[width=0.35\linewidth]{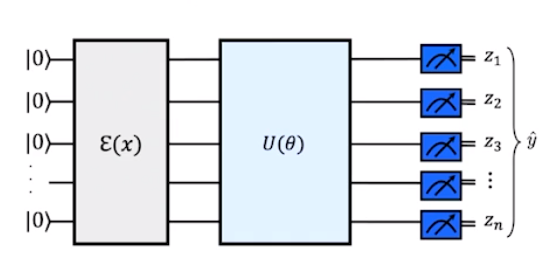}
\caption{Left: Quantum feature map using ZZFeatureMap, with superposition from Hadamard gates (red) and entanglement from CNOT gates (dark blue), contributing to quantum kernel construction. Right: Variational quantum circuit encoding input data $\epsilon(x)$, followed by trainable transformation $U(\theta)$ and measurement for class prediction.}
\label{fig:qml_circuits}
\end{figure}

Quantum machine learning (QML) offers a new paradigm by utilizing the principles of quantum mechanics to enhance traditional machine learning methods. Two key concepts in QML are quantum feature maps and quantum kernels \cite{havlicek2019supervised}, which are explained below:

(i) Quantum Feature Map: A quantum feature map encodes classical data into a quantum state by applying a sequence of quantum gates parameterized by the input features. This process effectively transforms the data into a high-dimensional space, using quantum properties such as superposition and entanglement. Superposition allows the system to represent multiple states at once, while entanglement introduces correlations between qubits that can capture complex relationships in the data. In Fig. \ref{fig:qml_circuits} left, the quantum feature map utilizes the Hadamard gates (red) to create superposition states, while entanglement between two qubits is introduced through the use of control-not gates (dark blue). By representing data as quantum states (quantum encoding), the quantum feature map can capture complex relationships that may be difficult for classical algorithms to identify.

(ii) Quantum Kernel: The quantum kernel measures the similarity between data points by computing the inner product of their quantum states, which are generated through the quantum feature map. In classical machine learning, kernels are used in methods like SVM to assess the similarity between data points in a transformed feature space. Similarly, a quantum kernel considers quantum states to potentially explore richer feature spaces, facilitating a more effective separation between classes. The formula for the quantum kernel is:

\begin{equation}
k(x_{i},x_{j})=|<\phi(x_{i})|\phi(x_{j})>|^{2}
\label{eq:quantum_kernel}
\end{equation}

where $\phi(x)$ is the quantum states produced by encoding the classical input data. In Fig. \ref{fig:qml_circuits} (right), the quantum feature map employs the ZZFeatureMap \cite{abraham2020qiskit}, followed by its reverse, creating an encoding and decoding process that captures correlations between qubits, which contributes to building the quantum kernel. The quantum kernel computation is used to train models such as QSVM, providing a new approach to kernel-based learning with quantum advantage.

We applied Principal Component Analysis (PCA) \cite{jolliffe2002principal} to reduce the kinematic variables of the jet dataset to six components, aligning the feature dimension with the six qubits in this study. This dimensionality reduction ensured computational efficiency while preserving the most relevant information, and it also helps reduce the required number of qubits, thereby mitigating the barren plateau problem \cite{mcclean2018barren} during optimization. The reduced feature set was then encoded into quantum states using a ZZFeatureMap with two repetitions, which leverages Hadamard gates and controlled-NOT gates to map classical data into a high-dimensional quantum feature space. For the quantum approach, we utilized the Qiskit Aer QASM simulator \cite{abraham2021qiskit} with 1024 shots to classify up and anti-up quark jets, with training performed on a local computer due to limitations in accessing IBM quantum servers. A trainable fidelity quantum kernel was constructed and optimized using the Simultaneous Perturbation Stochastic Approximation (SPSA) optimizer \cite{spall1992multivariate} with 100 iterations, enabling the Quantum Support Vector Machine \cite{havlicek2019supervised} to learn complex decision boundaries. The QML models were trained on the local computing clusters and required approximately one month to complete due to he high circuit simulation cost. This highlights the computational intensity of QML methods compared to classical models and reflects current hardware and simulation constraints.

In this analysis, we explore two QML models: the Quantum Support Vector Machine (QSVM) and the Variational Quantum Classifier (VQC) \cite{schuld2019evaluating}. The QSVM extends the classical SVM by using a quantum feature map to encode input data into a high-dimensional Hilbert space. Instead of relying on classical kernels, QSVM computes inner products between quantum states--forming a quantum kernel that captures complex similarities between data points. This approach can help the model learn more expressive decision boundaries, especially for problems involving high-dimensional or non-linearly separable data, where classical SVMs may fall short.

The VQC is a hybrid quantum-classical model that uses parameterized quantum circuits to perform classification tasks. In this approach, input data is first encoded into a quantum state using a quantum feature map. A variational circuit then processes the state using tunable parameters, which are optimized through classical methods to minimize a loss function. The circuit's output is measured and interpreted as a prediction, allowing the model to learn and adapt to the training data.

\begin{figure}[htbp]
\centering
\includegraphics[width=0.4\linewidth]{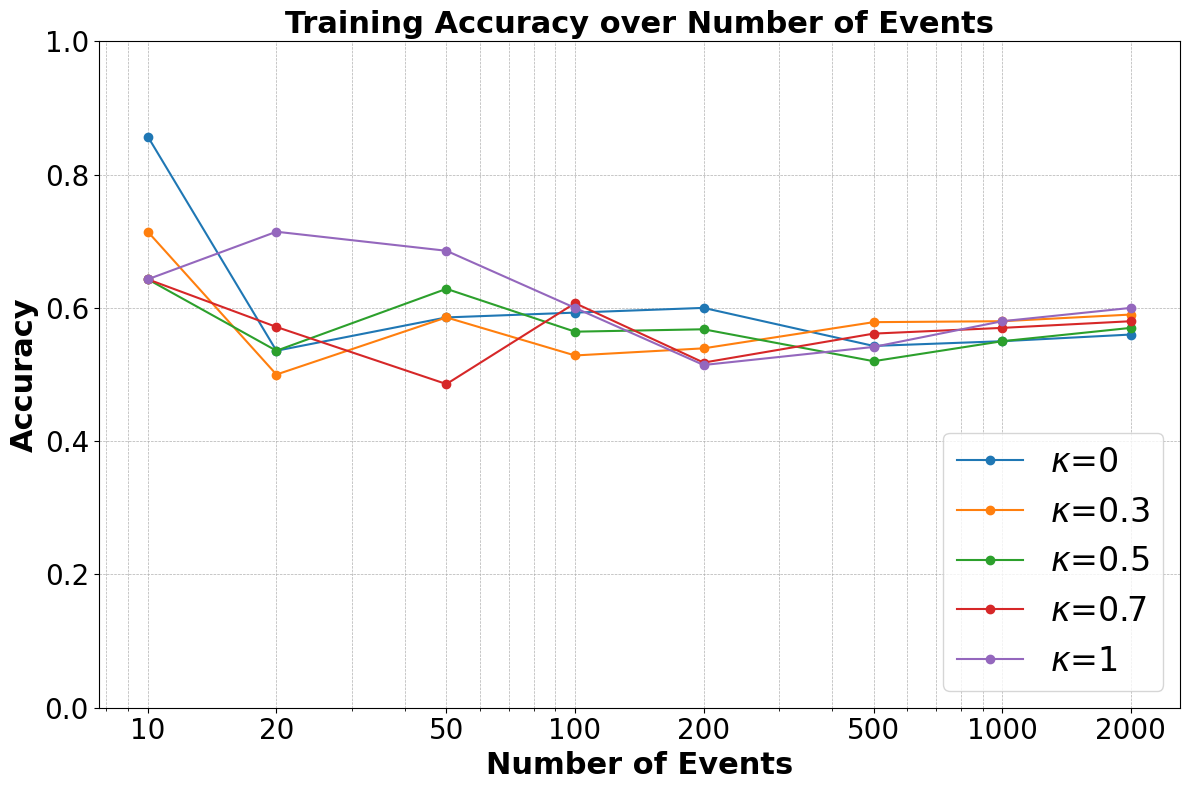}
\includegraphics[width=0.4\linewidth]{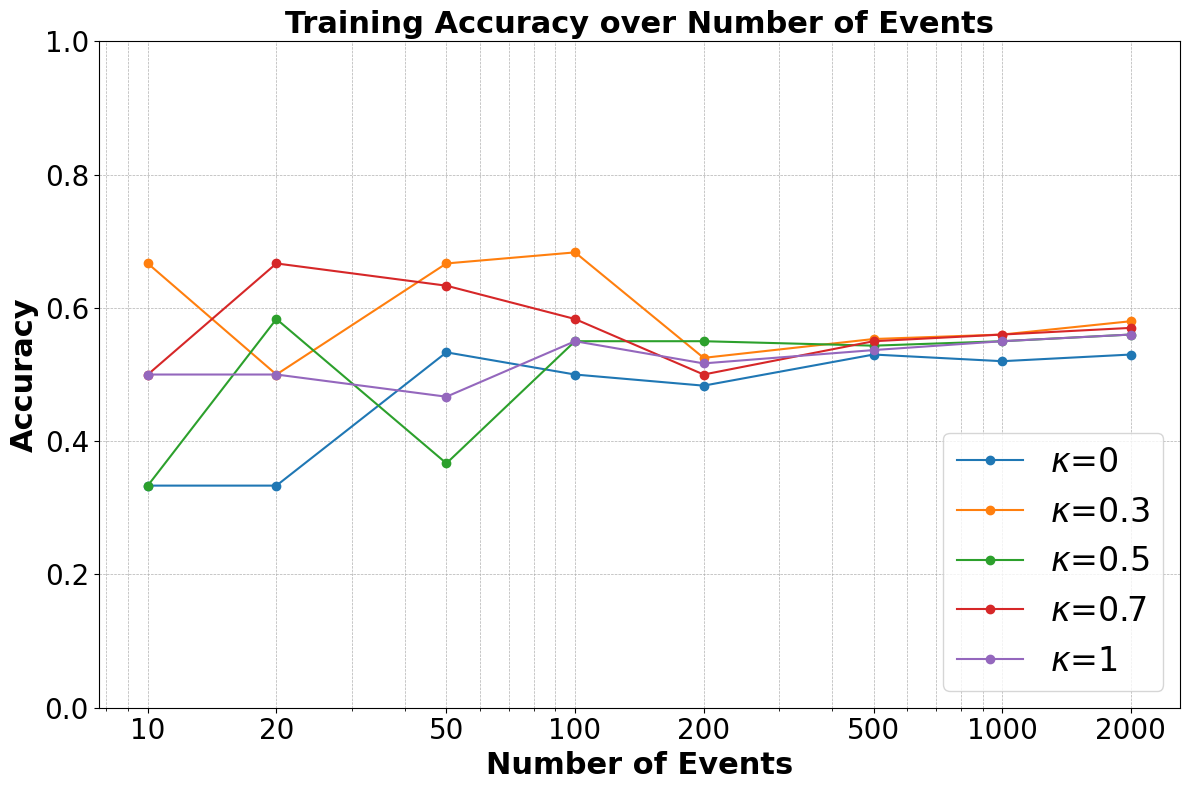}
\caption{Classification accuracy of VQC (left) and QSVM (right) models for up and anti-up quark jet discrimination, evaluated on a local Qiskit Aer QASM simulator for different training dataset sizes per class.}
\label{fig:qml_results}
\end{figure}

\begin{figure}[htbp]
\centering
\includegraphics[width=0.5\linewidth]{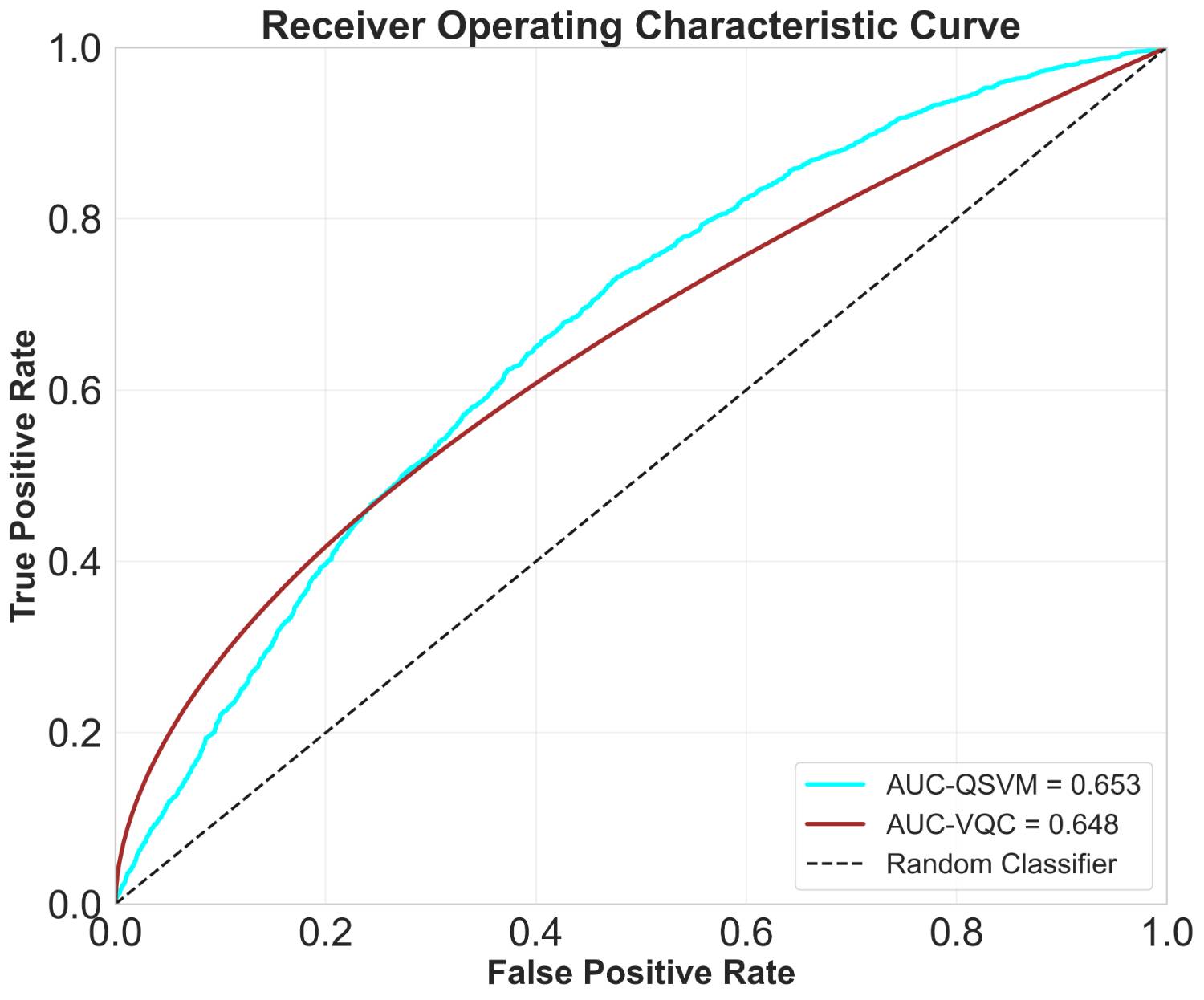}
\caption{ROC curves comparing VQC and QSVM performance on jet classification.}
\label{fig:quantum_roc}
\end{figure}

Fig. \ref{fig:qml_results} and Fig. \ref{fig:quantum_roc} show the classification accuracy and ROC curves for the VQC and QSVM models. Both quantum approaches achieve moderate performance in distinguishing up and anti-up quark jets, with AUC notably lower than those graph and attention-based models. This performance gap is particularly evident when compared to the GCN model as the most effective model in this study. The reduced AUC of the quantum models is likely due to the limited size of the training dataset (Due to the computational expense of training with over 2000 events), which constrained both circuit depth and generalization capability. We expect that with training with more events and improved quantum hardware availability, the performance of QML models could improve significantly, potentially closing the gap with classical deep learning methods. Our QML configuration (6 qubits, ZZFeatureMap with two repetitions, 1024 shots) led to markedly longer wall-clock times than classical baselines due to circuit simulation and kernel evaluation.

\subsection{Other Machine Learning Classifiers}

\begin{figure}[htbp]
\centering
\includegraphics[width=0.45\linewidth]{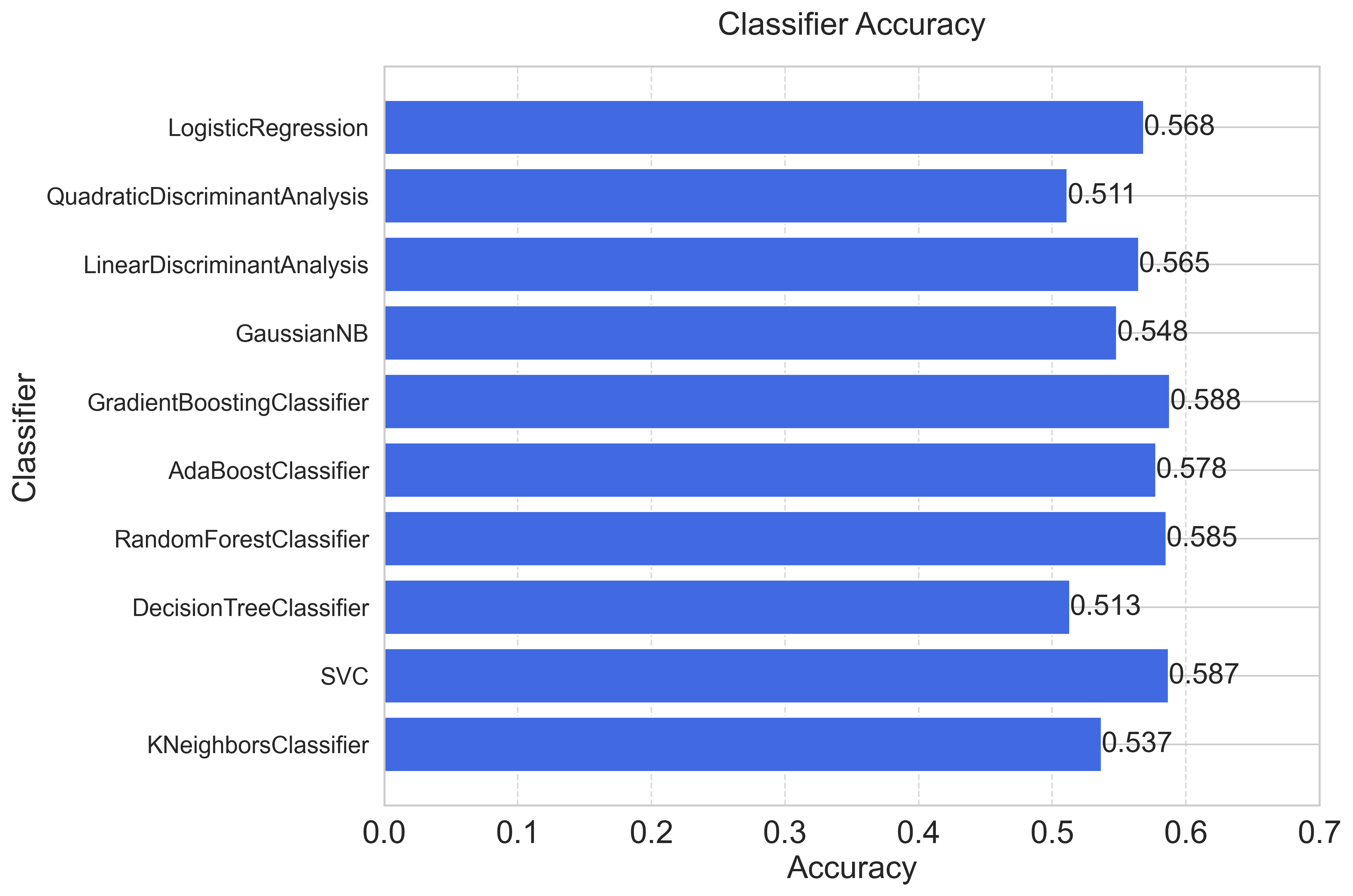}
\includegraphics[width=0.35\linewidth]{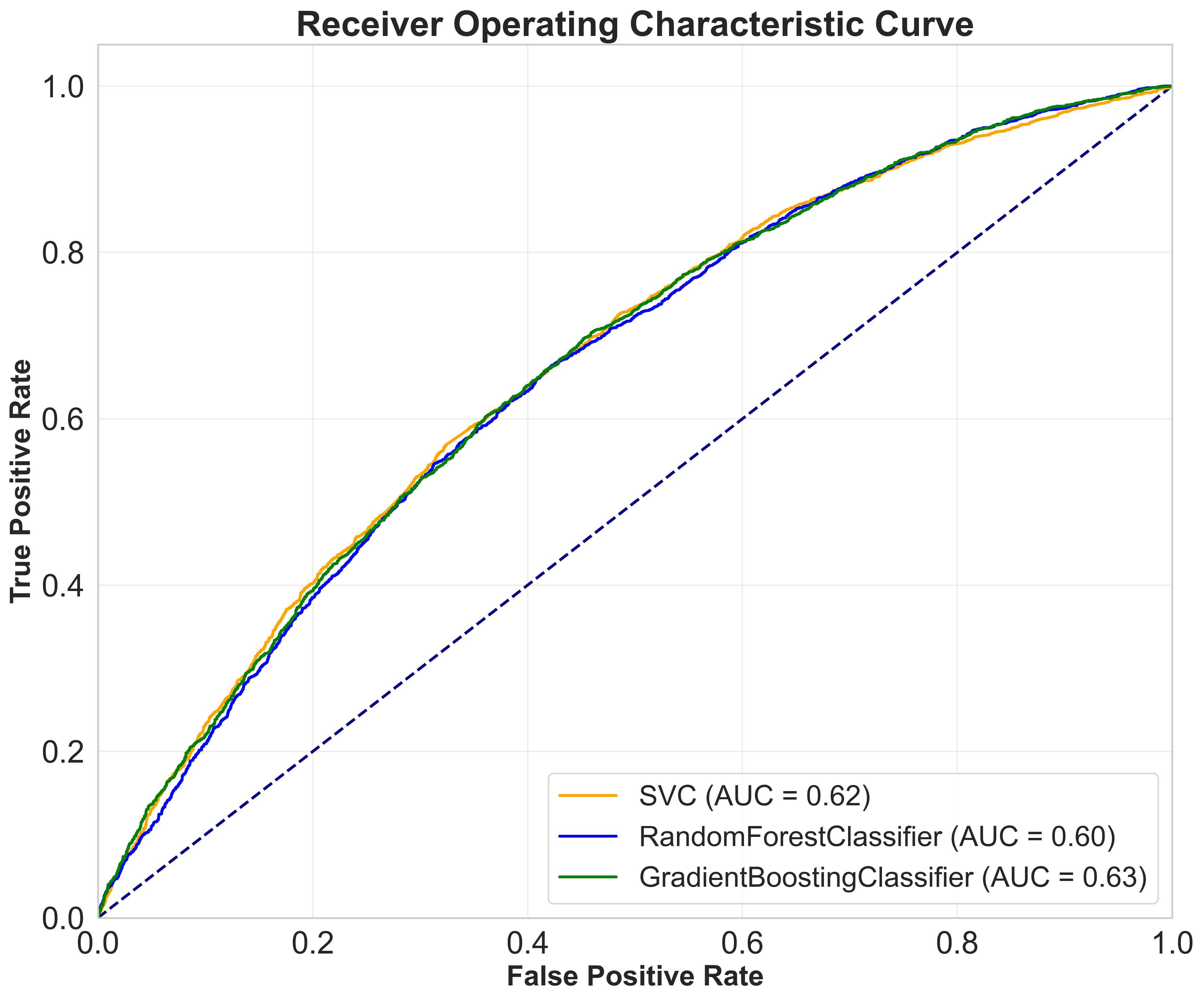}
\caption{Left: Accuracy of various ML classifiers for discriminating up and anti-up quark jets. Right: AUC scores for the three best-performing classical models: Support Vector Classifier, Random Forest, and Gradient Boosting.}
\label{fig:classical_ml}
\end{figure}

The Fig. \ref{fig:classical_ml} illustrates the accuracy and AUC of various classical machine learning classifiers in discriminating between up and anti-up quark jets using charge-related features. Among the tested algorithms, the Gradient Boosting \cite{friedman2001greedy} achieved the highest accuracy, slightly outperforming the Support Vector Classifier (SVC) \cite{cortes1995support} and Random Forest classifiers \cite{breiman2001random}, which are margin-based and ensemble methods known for their robustness and ability to capture nonlinear patterns. Logistic Regression, Linear Discriminant Analysis (LDA), and AdaBoost \cite{freund1997decision} also demonstrated competitive performance, suggesting that even relatively simple linear or shallow models can extract meaningful discriminative information from jet charge observables.

In contrast, the K-Nearest Neighbors (KNN) and Decision Tree classifiers performed less effectively, likely due to their sensitivity to high-dimensional feature spaces and potential overfitting to local fluctuations in the training data. The superior performance of margin-based methods (like SVC) and boosted ensembles (Gradient Boosting, AdaBoost) underscores their suitability for handling subtle differences in feature distributions, such as those present between up and anti-up quark jets.

\subsection{Extension to down anti-down quarks}

\begin{figure}[htbp]
\centering
\includegraphics[width=0.4\linewidth]{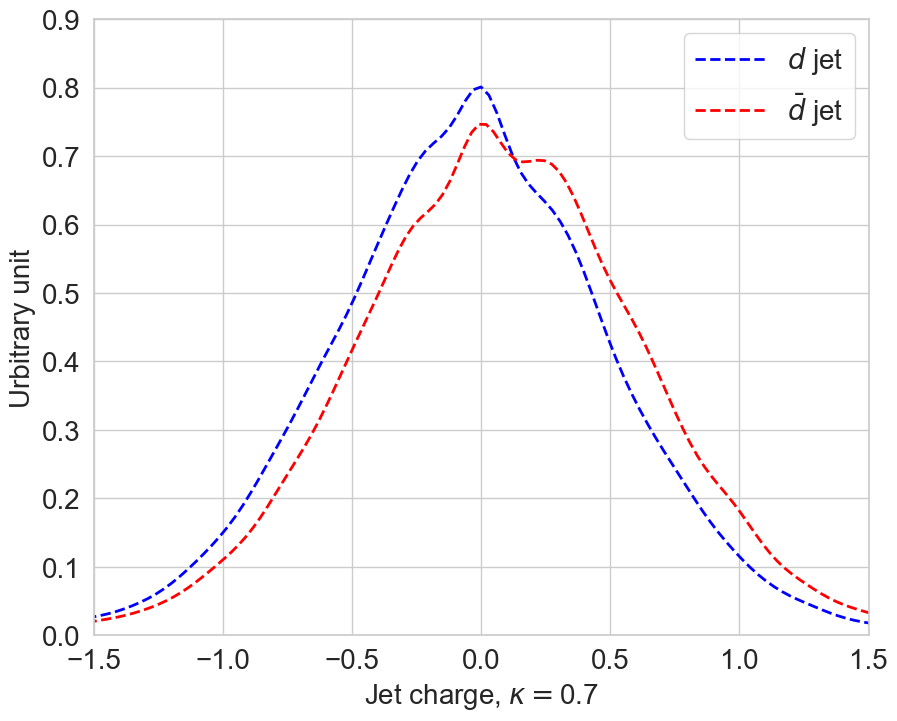}
\includegraphics[width=0.4\linewidth]{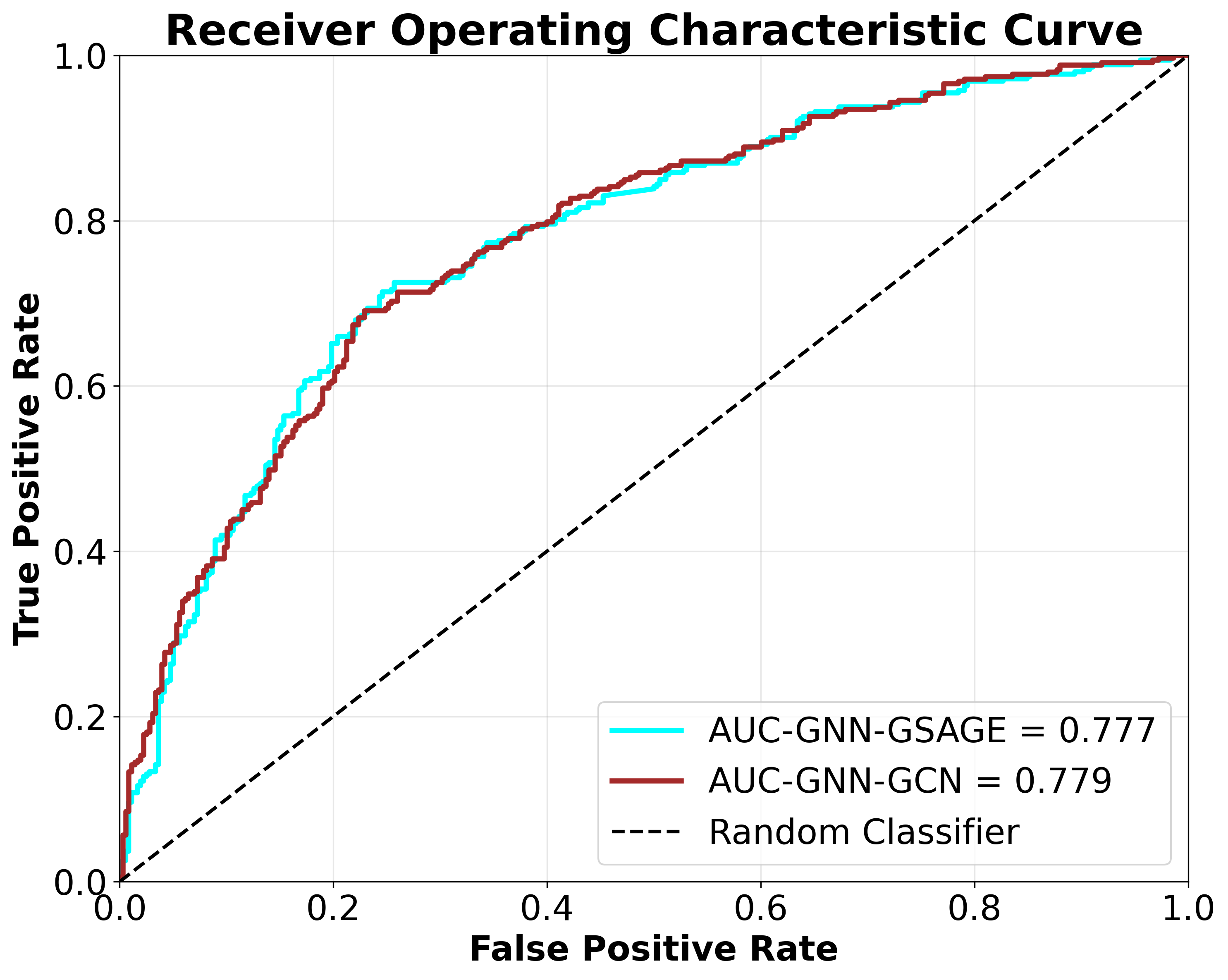}
\caption{Left: Distributions of $\mathcal{Q}_{j}$ for $d,\bar{d}$ jets obtained from $pp\to dg$ or $pp\rightarrow\bar{d}g$ events with $\kappa=0.7$. Right: ROC curves comparing GCN and GraphSAGE performance on $d\bar{d}$ jet classification.}
\label{fig:down_quarks}
\end{figure}

To extend the analysis to the discrimination of down (d) and anti-down ($\bar{\text{d}}$) quark-initiated jets, we applied the same GNN models (GCN and GraphSAGE), previously trained with up (u) and anti-up ($\bar{\text{u}}$) quark jets. The charge of the down quark is $-\frac{1}{3}$, compared to the up quark's charge of $\frac{2}{3}$, resulting in a smaller charge difference between d and $\bar{\text{d}}$ quarks. This reduced charge difference makes the discrimination task more challenging, as the jet charge distributions, exhibit greater overlap for d and $\bar{\text{d}}$ jets compared to u and $\bar{\text{u}}$ jets as shown in Fig. \ref{fig:down_quarks} left. For the down and anti-down quark jet analysis, a separate dataset of 35,000 simulated events was used. The GNN models, exploring the relational structure of jet constituents, achieved an AUC of 0.77 for d-$\bar{\text{d}}$ discrimination, slightly lower than the AUC of 0.883 for u-$\bar{\text{u}}$ discrimination with GCN (Fig. \ref{fig:down_quarks} right). This reduction in performance is attributed to the smaller charge difference, which diminishes the distinguishing features in the jet charge distributions, making it harder for the GNN to capture complex patterns effectively. These results highlight the robustness of GNNs in handling complex jet structures, even for more challenging discrimination tasks, and suggest potential for further optimization to improve performance in such scenarios.

\section{Results and Discussion}
\label{sec5}

This study explored several machine learning models, both classical and deep learning-based for discriminating between up and anti-up quark jets using jet charge information. Among the classical models, support Vector Classifier (SVC) and Gradient Boosting achieved the highest performance. Most notably, the best results were achieved using Graph Neural Network (GNN) models, which leveraged the particle-level structure of jets. The Graph Convolutional Network (GCN) architecture achieved the highest performance with an AUC of 0.883. A comparison of AUC and accuracy across all models is provided in Table \ref{tab:performance_comparison} (each model except QML, was executed 10 times to calculate the average accuracy-AUC and its standard deviation), highlighting the relative strengths of each approach. These findings demonstrate that GNNs are particularly well-suited for jet charge discrimination tasks, offering a powerful alternative to traditional and image-based models.

To directly compare our machine learning models with the traditional jet charge approach, we evaluated the standard observable $Q_{j}(\kappa)$ (Eq. \ref{eq1}) across the same set of $\kappa$ values. We found that the best-performing choice, $\kappa=0.7$, yields an AUC of 0.60 for discriminating $u$ and $\bar{u}$ jets. This comparison makes clear that, while the classical jet charge observable provides modest separation power, the machine learning models--particularly the CNNs and GNNs--achieve substantially higher AUC values, thereby quantifying the improvement and highlighting the necessity of ML techniques for jet charge discrimination.

\begin{table}[htbp]
\centering
\caption{Performance comparison of different models for $u$-$\bar{u}$ jet classification in terms of AUC and accuracy.}
\begin{tabular}{l c c}
\hline
\textbf{Model} & \textbf{AUC} & \textbf{Accuracy} \\
\hline
Traditional $Q_{j}(\kappa=0.7)$ & $0.600\,\pm\,0.005$ & $0.572\,\pm\,0.005$ \\
Deep Neural Network (DNN) & $0.674\,\pm\,0.002$ & $0.628\,\pm\,0.004$ \\
Baseline CNN & $0.667\,\pm\,0.003$ & $0.621\,\pm\,0.003$ \\
Optimized CNN & $0.717\,\pm\,0.003$ & $0.709\,\pm\,0.004$ \\
Graph Conv. Network (GCN) & $0.883\,\pm\,0.002$ & $0.802\,\pm\,0.002$ \\
GraphSAGE Network & $0.860\,\pm\,0.003$ & $0.783\,\pm\,0.003$ \\
Transformer & $0.865\,\pm\,0.005$ & $0.785\,\pm\,0.005$ \\
Vision Transformer (ViT) & $0.756\,\pm\,0.007$ & $0.723\,\pm\,0.006$ \\
QSVM & $0.653\,\pm\,0.004$ & $0.611\,\pm\,0.005$ \\
VQC & $0.648\,\pm\,0.004$ & $0.596\,\pm\,0.005$ \\
\hline
\end{tabular}
\label{tab:performance_comparison}
\end{table}

\begin{figure}[htbp]
\centering
\includegraphics[width=0.6\linewidth]{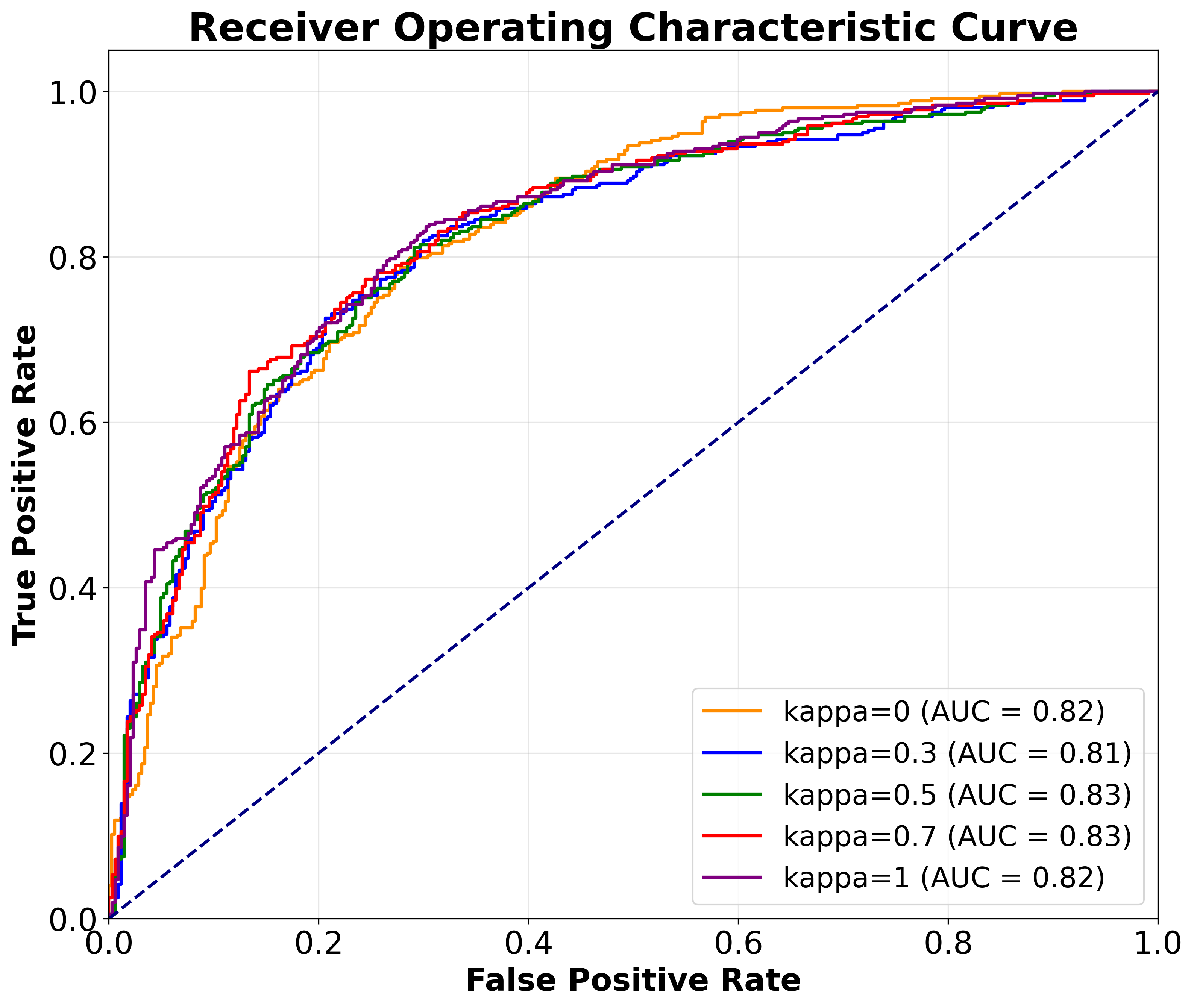}
\caption{ROC curves showing the performance of GCN models trained with different $\kappa$ parameters in the jet charge calculation.}
\label{fig:kappa_sensitivity}
\end{figure}

To explore the sensitivity of the GCN model to the choice of the jet charge weighting 
parameter $\kappa$, we trained separate GCN models, each receiving only the jet charge 
observables $Q_{1,\kappa}$ and $Q_{2,\kappa}$ computed at a single $\kappa$ value as 
input, in contrast to the main comparison in Table \ref{tab:performance_comparison} where all 
$\kappa$ values are provided simultaneously. The results, shown in 
Fig. \ref{fig:kappa_sensitivity}, indicate that the model performs consistently well, with AUC 
scores ranging from 0.81 to 0.83, slightly below the AUC of $0.883$ achieved in the 
main comparison, a difference attributed to the reduced input information available to 
the network when restricted to a single weighting scale. Among the tested values, 
$\kappa = 0.5$ and $\kappa = 0.7$ yielded the best discrimination between up and 
anti-up quark jets. This stability across different $\kappa$ settings suggests that 
the GCN is not only effective in capturing charge-related patterns, but also robust 
to the specific weighting of the input features.

To study detector effects, all samples were generated using MadGraph5\_aMC@NLO and PYTHIA8, are subsequently processed through Delphes3, employing the high-luminosity CMS detector configuration to simulate realistic detector effects and to reconstruct the final-state objects. The reconstructed charged-particle tracks are used as jet constituents, replacing the particle-level inputs used in the previous sections. The GCN model, which achieved the best performance at the particle level, was retrained on the detector-level samples using the same architecture and hyperparameters. For $u\bar{u}$ discrimination, the detector-level GCN achieves an AUC of 0.812, compared to 0.883 at the particle level, representing a degradation of approximately 0.07 AUC units. For $d\bar{d}$ events, the detector-level AUC is 0.685, compared to 0.779 at the particle level, reflecting a degradation of 0.09. This reduction is expected and is primarily attributed to charged-particle reconstruction inefficiencies, and loss of neutral particles. the GCN continues to substantially outperform the traditional momentum-weighted jet charge observable at the particle-level for both quark flavors.

The ability to discriminate between up and anti-up quark jets using machine learning, is particularly relevant for LHC analyses that rely on identifying the electric charge of hadronic jets. While this study shows promising results for quark jet charge discrimination, there are still some important limitations to consider. One of the main ones is that the analysis was performed entirely on simulated data. To move closer to real-world applications at the LHC, future work should explore how well these models perform on reconstructed or detector-level data, where effects like resolution, pile-up, and detector noise can impact the charge information. Another limitation is the relatively small size of the training dataset. Although the models performed well especially the GNNs, larger datasets would likely help improve generalization and stability, particularly for more complex architectures. This also opens the door to expand the current binary classification task (e.g., up vs. anti-up) to more challenging scenarios like multiclass classification across different quark flavors, or even including gluon jets. In addition, while we explored classical and graph-based models in detail, we were unable to run the QML models, such as QSVM and VQC, on IBM's quantum servers due to hardware access limitations. As a result, their evaluation remains incomplete in this version of the study and should be revisited in future work. This constraint also limited the size of the training dataset for QML models; with access to better quantum hardware or simulators, we expect that training on larger event samples could lead to significant performance improvements.

\section{Summary and Conclusions}
\label{sec6}

The determination of jet charge is an important ingredient for a variety of measurements and searches at the LHC, ranging from charge asymmetries in $W$ boson decays and top-quark property studies to the identification of final states in extensions of the Standard Model. Traditional observables, such as the $p_{\rm T}$-weighted jet charge, provide only modest separation between quark and antiquark jets.

In this work, we investigated the potential of modern machine-learning techniques, with a particular focus on graph-based architectures, to enhance jet charge discrimination. Using a controlled simulation framework with $u$ and $\bar{u}$ jets, we systematically compared classical methods, convolutional networks, and graph neural networks. The best-performing model, a Graph Convolutional Network, achieved an AUC of $0.883\,\pm\,0.002$, substantially outperforming the conventional jet charge observable and demonstrating the ability of graph-based approaches to exploit the fine-grained particle-level structure of jets.

Although this study was performed at the particle-level using simulated data, the results establish a robust methodological foundation that can be directly transferred to experimental analyses. Future work will focus on validating these techniques at the detector level, where pile-up, finite resolution, and reconstruction efficiencies must be taken into account. Further directions include exploring larger training datasets, more complex architectures such as quantum machine-learning models, and extensions beyond binary classification to include multiple quark flavors and gluons.

In conclusion, the application of modern machine-learning techniques, and graph-based networks in particular, provides a powerful and flexible framework for jet charge determination. These methods show clear advantages over traditional observables and are mature enough to be incorporated into upcoming detector-level studies and LHC analyses.

\section*{Acknowledgments}
The authors would like to sincerely thank Dr. Gh. Haghighat for his insightful comments and constructive feedback, which significantly improved the quality of this manuscript. We are also grateful to the Institute for Research in Fundamental Sciences (IPM) for providing the computational resources and server access that made this research possible.

\section*{Data availability}
The data and code supporting the findings of this study are publicly available \cite{zenodo2024jet}.

\end{document}